\documentclass[a4paper,11pt]{article}
\pdfoutput=1
\usepackage{jinstpub}
\usepackage{array}
\usepackage{subcaption}

\newcolumntype{A}{>{\raggedright\arraybackslash}m{9cm}}

\title{A Model for the Global Quantum Efficiency for a TPB-based Wavelength-Shifting 
System used with Photomultiplier Tubes in Liquid Argon in MicroBooNE}
\author[a,1]{S.F. Pate,\note{Corresponding Author}}
\author[b]{T. Wester,}
\author[b]{L. Bugel,}
\author[b]{J. Conrad,}
\author[a]{E. Henderson,}
\author[b,c]{B.J.P. Jones,}
\author[a]{A.I.L.~McLean,}
\author[b]{J.S. Moon,}
\author[b,d]{M. Toups,}
\author[b]{and T. Wongjirad} 
\affiliation[a]{Physics Department, New Mexico State University, Las Cruces, NM, 88003, USA}
\affiliation[b]{Physics Department, Massachusetts Institute of Technology, Cambridge, MA, 02139, USA}
\affiliation[c]{Physics Department, University of Texas at Arlington, Arlington, TX, 76019, USA}
\affiliation[d]{Fermi National Accelerator Laboratory, Batavia, IL, 60510, USA}
\emailAdd{pate@nmsu.edu}

\abstract{
We present a model for the Global Quantum Efficiency (GQE) of the MicroBooNE optical units. An optical unit consists of a flat, circular acrylic plate, coated with tetraphenyl butadiene (TPB), positioned near the photocathode of a 20.2-cm diameter photomultiplier tube. The plate converts the ultra-violet scintillation photons from liquid argon into visible-spectrum photons to which the cryogenic phototubes are sensitive. The GQE is the convolution of the efficiency of the plates that convert the 128 nm scintillation light from liquid argon to visible light, the efficiency of the shifted light to reach the photocathode, and the efficiency of the cryogenic photomultiplier tube. We develop a GEANT4-based model of the optical unit, based on first principles, and obtain the range of probable values for the expected number of detected photoelectrons ($N_{\rm PE}$) given the known systematic errors on the simulation parameters. We compare results from four measurements of the $N_{\rm PE}$ determined using alpha-particle sources placed at two distances from a TPB-coated plate in a liquid argon cryostat test stand. We also directly measured the radial dependence of the quantum efficiency, and find that this has the same shape as predicted by our model. Our model results in a GQE of $0.0055\pm0.0009$ for the MicroBooNE optical units. While the information shown here is MicroBooNE specific, the approach to the model and the collection of simulation parameters will be widely applicable to many liquid-argon-based light collection systems.} 


\begin{document}
\maketitle

\section{Introduction}

Light-collection systems in liquid-argon time projection chambers (LArTPCs) perform critical functions that are complementary to the data collected by the TPC system.  The fast (few nanosecond) response timing of the photomultipliers (PMTs) used in these systems allows for a powerful triggering capability, critical to reduction of data volumes that would otherwise be required in a pure minimum-bias operation. A triggering capability is very important for rare events, like supernova or dark matter observations. If the PMTs view the entire active volume in a uniform fashion, then the light collection system is valuable for calorimetry, and the fast time response can also be used for correlation of vertex location with optical energy; these two capabilities are important for distinguishing signal events from background events.  In addition, a fast and efficient light-collection system enables the rejection of cosmogenic backgrounds.

Because the scintillation light from liquid argon is in the vacuum ultra-violet (VUV), at 128 nm, it is completely absorbed by the glass enclosure of a PMT.  A wavelength-shifting system is needed, to bring the light into a wavelength range suitable for available PMTs.  In this paper, we describe a system employing a layer of tetraphenyl butadiene (TPB) applied to an acrylic plate positioned above the photocathode surface.  TPB is well-known to be highly absorbing of the VUV photons from liquid argon scintillation, reemitting photons in the visible spectrum with a distribution peaked at 420 nm \cite{Gehman2011116}.  In this arrangement, the TPB surface on the acrylic plate is the active surface of the light-collection system.

The MicroBooNE detector comprises an 89-ton active volume of liquid argon, with a system of 32 PMTs arranged in a single plane behind the TPC collection wires~\cite{Conrad-JINST-10-06-T06001}.  Each 14-stage 20.2-cm diameter Hamamatsu R5912-02mod PMT is surrounded by a cylindrical magnetic shield aligned with the symmetry axis of the PMT, and a TPB-covered plate of radius $R=15.2$ cm is positioned normal to the symmetry axis and 0.64 cm above the photocathode~\cite{Briese-JINST-8-07-T07005}; this assembly of phototube, magnetic shield and TPB-coated plate will be referred to henceforth as an ``optical unit.'' To understand the performance of the MicroBooNE light-collection system for use as a trigger and calorimeter, we must determine the probability (the quantum efficiency) that a 128-nm photon impinging on the TPB-coated plate of an optical unit will produce a photoelectron in the unit's PMT.  

An important component of the understanding of the optical unit is the dependence of the quantum efficiency on the location of incidence of the photon on the TPB-coated plate, as a function of radial coordinate $r$ and azimuthal coordinate $\phi$. This quantity we will call the differential quantum efficiency, $Q(r,\phi)$. We studied $Q(r,\phi)$ via two experimental measurements:

\begin{itemize}
	\item A measurement of the absolute response of an optical unit, using alpha-emitting sources to create point-like sources of scintillation photons in liquid argon.
	\item A measurement of the position-dependence of the efficiency, performed by exposing a set of points on the TPB surface to UV light from an optical fiber.
\end{itemize}
Based on measurements of the gain of large PMTs as a function of photocathode position~\cite{Koblesky201240,Abbasi2010139}, we do not expect a strong azimuthal angle dependence of the quantum efficiency arising from the PMT construction itself.  In~\cite{Koblesky201240}, a study of a nearly identical PMT (the Hamamatsu R5912), figure 4b suggests the azimuthal variation is not more than $\pm 10$\% at a given radius from the center.  Additionally, the magnetic shield removes the influence of the earth's magnetic field \cite{Acciarri:2016smi} which might have caused an azimuthal dependence in the efficiency.  The only remaining possible source of azimuthal dependence would be gaps in the TPB coverage on the plate, which we also do not expect \cite{Acciarri:2016smi}.  With these facts in mind we will use a simple mathematical model for the differential quantum efficiency:
\begin{equation}
Q(r,\phi) = f(r)[1+\delta\sin\phi].
\label{dqe_def}
\end{equation}
In our measurements we will try to set limits on the size of $\delta$ and establish the functional form for $f(r)$.

This form for the efficiency is very useful when performing detailed simulations of the light collection process. In the interpretation of real data, on the other hand, we do not know where individual photons have been incident on the TPB plate; in this case it is more useful to consider a global quantum efficiency $\left<Q\right>$
which is the average of the differential efficiency over the surface of
the plate:
\begin{equation}
\left<Q\right> = \frac{1}{\pi R^2}\int_0^R dr \int _0^{2\pi} d\phi \, rf(r)[1+\delta\sin\phi] = \frac{2}{R^2} \int_0^R dr\, r\,f(r)
\label{dqe_int}
\end{equation}
where $R$ is the radius of the TPB plate.  Note that the azimuthal variation parameter $\delta$ disappears in the integration.   We call this the global quantum efficiency (GQE) because this represents the efficiency of all elements of the system working together, not just that of the PMT.

Using the TallBo test stand at Fermi National Accelerator Laboratory we have made four measurements of the average number of photoelectrons ($N_{\rm PE}$) observed in the PMT when using an alpha source in liquid argon to illuminate the optical unit.  The alpha source was located at two different positions above the center of a MicroBooNE optical unit in liquid argon. We have created a simulation of this test stand which incorporates the alpha scintillation in the liquid argon, the propagation of scintillation photons to the TPB plate, and the re-emission of visible photons within a thin layer of TPB, and tracks these to a simulated PMT to obtain a simulated value for $N_{\rm PE}$.
We will show that the results of the TallBo measurements are within the range of $N_{\rm PE}$ values predicted by our simulation.  

In addition, in a separate apparatus, we have directly measured the position dependence of the optical response by illuminating specific points on the surface of the TPB-coated plate
using optical fibers.  We will show that the simulation reproduces the shape 
of the radial dependence.

Section~\ref{sec:absnorm} will describe the four measurements of the $N_{\rm PE}$ that we will use to cross check our simulation. This allows an introduction to the apparatus we will be modeling. Section~\ref{sec:posdep} describes our measurements of the position dependence of the light response.   Section~\ref{sec:sim} will discuss the simulation and its parameters.  Section~\ref{results} compares the $N_{\rm PE}$ measurement, and the radial dependence of the optical response, to that from our simulation, and describes the GQE that we determine for a MicroBooNE optical unit.  We will note that  although the simulation is MicroBooNE-specific, the general approach is widely applicable, and should allow successful determination of the GQE in other liquid-argon-based light collection systems, such as the flat-panel systems under consideration for DUNE~\cite{wunderbars}.

\section{Description of Apparatus and Measurement of $N_{\rm PE}$\label{sec:absnorm}}

The apparatus used for the absolute normalization of the quantum efficiency is described in great detail in reference~\cite{Jones-JINST-8-07-P07011}; a summary is provided here.  A polonium-210 alpha source was positioned at two different distances ($D = 20.3$ cm and 36.8 cm) above a standard MicroBooNE optical unit, and both placed within the ``TallBo'' cryostat at the cyrogenic test facility at Fermi National Accelerator Laboratory. The cryostat was filled with very high purity liquid argon (nitrogen $<1$ ppm, oxygen $<100$ ppb, water $<100$ ppb). The 5.3-MeV alpha particles emitted by the source are stopped in the liquid argon after traveling less than 1 mm; the kinetic energy of the alpha particle ionizes some argon atoms, leading to the formation of excimers and subsequent scintillation in the ultra-violet (128 nm).  Some of the VUV photons from the scintillation reach the TPB plate and are absorbed and re-emitted near 420 nm; some of these visible photons will reach the PMT photocathode and produce a photoelectron, initiating the production of a signal at the PMT anode.  In~\cite{Jones-JINST-8-07-P07011}, the anode signal was measured as a function of the distance of the source from the center of the TPB plate, and also as a function of nitrogen contamination. Measurement of the distance and nitrogen contamination dependencies allowed a determination of the attenuation length for the absorption of 128 nm light by nitrogen dissolved in liquid argon.  A very significant result of~\cite{Jones-JINST-8-07-P07011} is that for the very pure argon described above the attenuation length is in excess of 30 meters.  Therefore we can use data from that study taken with the purest available liquid argon to determine the absolute normalization of the quantum efficiency, because in that case we may ignore attenuation altogether.

The PMT was supplied with HV of 1100 V, which produces a stable gain of $1.00\times10^{7}$. An optical fiber pointed at the photocathode can be used to artificially illuminate the PMT with visible light from pulsed LEDs. The gain was measured several times around the time of the study by counting the area of single photoelectron pulses generated by flashing the LED at low intensity, and found to be stable to within 2\%.

Signal and high voltage from the PMT are carried by a single cable, and outside the cryostat the DC high voltage component is split from the AC signal component by a HV splitter unit. For this study we used a metal-cased splitter unit which is similar in layout to the splitters for MicroBooNE.

The signal component from the splitter is fed via RG58 cable to a Tektronix DPO5000 oscilloscope, which is terminated at 50 $\Omega$. This oscilloscope is used to produce histograms of pulse amplitudes or areas within a chosen time window around a trigger. For this measurement we histogram the area of the first 50 ns of PMT pulses which pass a 30 mV trigger. This threshold is about 2.5 times the amplitude of an average single-photoelectron (SPE) pulse, sufficiently low to capture the majority
of alpha-particle-initiated pulses.  

\begin{figure}[t]
\begin{center}
\includegraphics[scale=0.6]{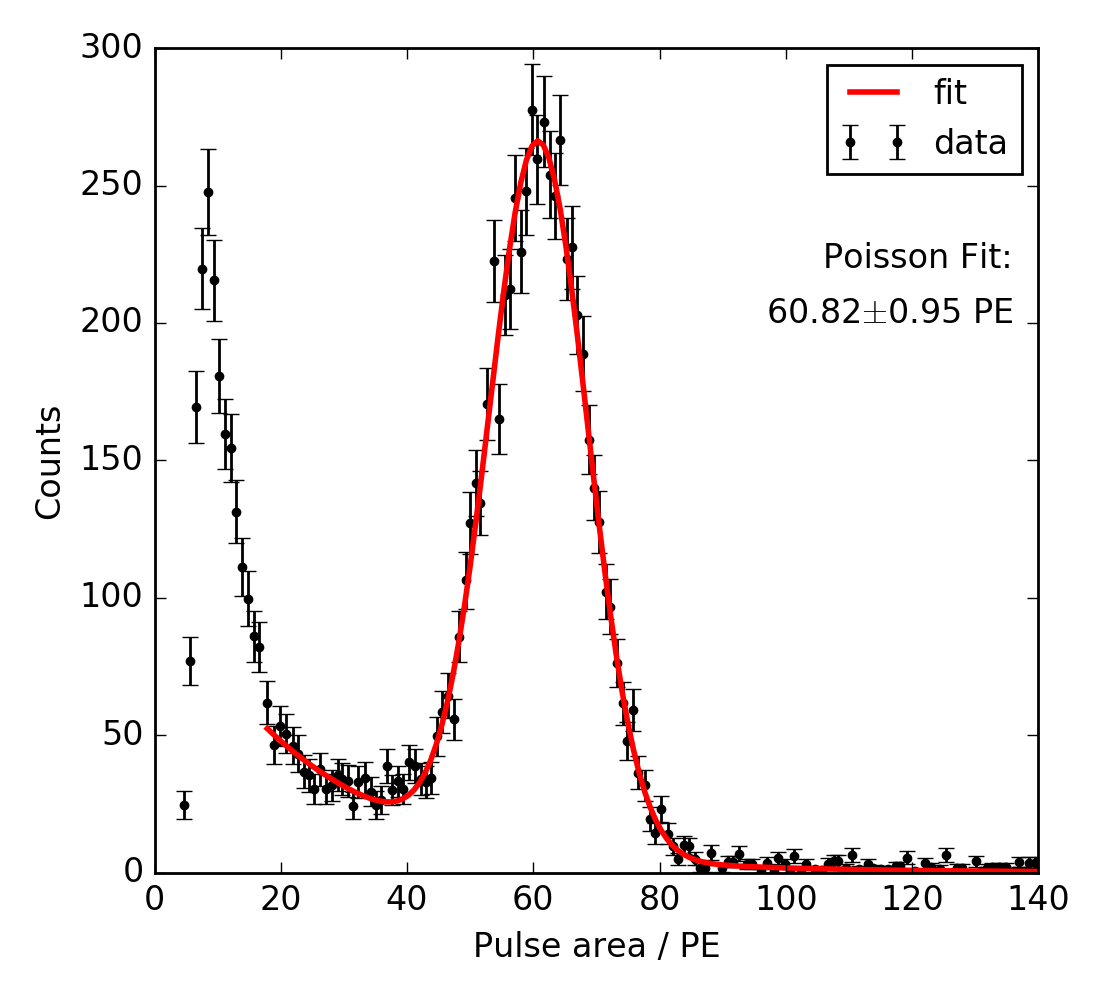}
\end{center}
\caption{Example of data taken at the near ($D=20.3$ cm) location, called ``Near-measurement 1.''  The fit used to extract the mean number of PE, which is used in the analysis, is also shown.   Details of the fit are explained in the text.}
\label{prettyPEplot}
\end{figure}

Data were taken with the source placed at $D=20.3$ and $36.8$ cm away from the face of the TPB plate, henceforth called the near and far positions. Fig.~\ref{prettyPEplot} shows an example of a data set taken at the near position as a function of the measured counts, called $x$ below.   As discussed in \cite{methanepaper},  the signal is overlaid on a background that is well described by an exponential plus an overall constant offset.     In this analysis, we make a small departure from the approach of Ref.~\cite{methanepaper}.  In that paper, a modified Poisson function was used, corrected for source occlusion and normalized to an independently measured SPE scale.  In this work, on the other hand, we assume a Poissonian form and extract the SPE scale shift from the fit directly, without an independent constraint.  The SPE scale shift extracted here is thus not a physically meaningful parameter in itself, but represents the deviation from an initial and approximate calibration scale that is not expected to be equivalent between runs.  Unlike the SPE parameter of Ref.~\cite{methanepaper}, therefore, it should not be interpreted as a proxy for the PMT gain, but rather as an independent nuisance parameter in fits that extract $\mu$.  The form of the fit used is then:\\
\begin{equation}
\frac{N}{\sqrt{2\pi}\mu}\exp[-(x-\mu)^2/2\mu S] + A\exp(-x/B)+C,
\end{equation}
where $N$ is the normalization of the $\alpha$-source rate, $\mu$ is the mean number of PE ($N_{\rm PE}$),  $S$ is the SPE scale shift, and $A$, $B$, and $C$ are constants that describe the background.  An example fit is shown in. Fig.~\ref{prettyPEplot}.     

\begin{table}
\begin{center}
\begin{tabular}{ ccc }
Data Sample &  $N_{\rm PE}$ & SPE Scale Shift \\
\hline
  Far-measurement 1	& $37.5 \pm 0.5$ 	& $1.04 	\pm 0.02$ \\
  Far-measurement 2	& $30.9 \pm 0.5$ 	& $0.95 	 \pm 0.02$ \\
  Near-measurement 1 	& $60.8 	\pm 1.0$ &	$1.12 \pm 0.02$ \\
   Near-measurement 2 	& $62.0 	\pm 0.7$ & $1.20 \pm 0.02$ \\
\end{tabular}
\end{center}
\caption{Four measurements of the $N_{\rm PE}$ that will be used to validate the
simulation (to be presented in section~\ref{sec:sim}).   Column 2 presents the $N_{\rm PE}$ and the error from the fit on 
this parameter.   Column 3 presents another fit variable, the
SPE scale shift, and its error.  These quantities are described in the text. \label{NPESPE}}
\end{table}

From Table~\ref{NPESPE}, one can see that the the two far measurements disagree.  The average of these two $N_{\rm PE}$ measurements is $34.2\pm 0.7$ photoelectrons and the standard deviation is 3.3 photoelectrons.   Thus, the average is 4.5$\sigma$, or a 9\% deviation, from either measurement.      This indicates that the two far measurements suffered systematic effects that caused them to differ beyond statistical error.

Our simulation, to be discussed in section~\ref{sec:sim},
introduces Gaussian priors on the input parameters that encompass possible systematic effects that could affect these measurements, including those that would result from different levels of impurities (hence a difference in the number of photons from the alpha source that can reach the plate), and differences in PMT response due to temperature, which changes within the tube as it cools.    Capability to address this spread is an essential feature of the simulation, as discussed in section~\ref{results}.

\section{Position Dependence of the Quantum Efficiency}\label{sec:posdep}

\begin{figure}[ht]
\begin{center}
\includegraphics[scale=0.5]{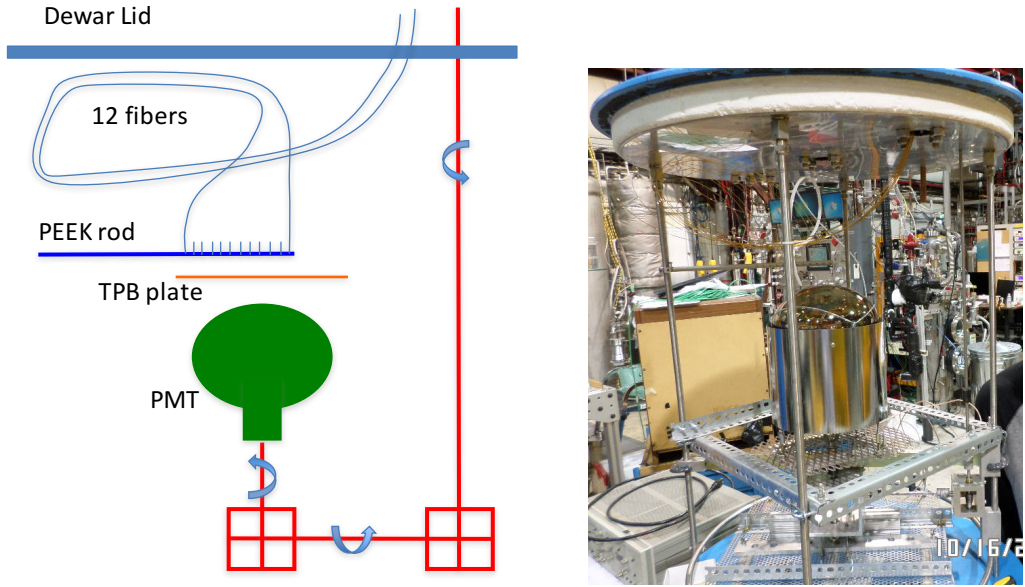}
\end{center}
\caption{Schematic diagram, and photograph, 
of the apparatus for the measurement of the position-dependence of the quantum
efficiency.  12 optical fibers were held in place by a PEEK rod over the TPB-coated plate.  
The optical unit (TPB plate plus the PMT and magnetic shield) could be
rotated underneath the fixed fibers by means of a system of shafts and gear boxes.}
\label{apparatus_figure}
\end{figure}

We created an apparatus 
(see figure~\ref{apparatus_figure}) to hold a standard MicroBooNE optical unit 
(including the PMT itself, a resistor-chain base soldered to the PMT leads, a magnetic shield, and a TPB-coated 
acrylic plate) in a 300-liter cryogenic dewar called ``Stella''.
The optical unit was oriented with its symmetry axis vertical, 
with the TPB plate on top.  Above the TPB plate, a PEEK rod was positioned
to pass over a diameter of the plate.  This PEEK rod held in place a set of 12 quartz optical fibers, spaced
1.27 cm apart, with the 2nd of the 12 fibers positioned directly over the center of the plate.  
The end of each fiber was approximately 8 mm from the TPB plate.  These fibers were 
approximately 110 cm long; the other end of each fiber lay outside the dewar, so that they could be illuminated 
with ultraviolet light from a light-emitting diode (LED).  
In addition, a set of rods and gears allowed a rotation
of the optical unit about its symmetry axis, underneath the stationary PEEK rod.  
By illuminating individual fibers, and
by rotating the optical unit, we were able to survey the TPB plate both radially and azimuthally.

The PMT was operated using a Spectrum Techniques UCS 30 unit~\cite{spectechUCS30}, 
which supplied high-voltage and also observed the 
anode signal via the same AC-coupled circuit described above.  
The anode signal pulse was amplified and digitized within the UCS 30 unit.
A histogram from the UCS 30 was viewable on a computer via a USB connection.  We also examined individual anode 
signal pulses using an oscilloscope.

We used optical fibers from Molex Polymicro~\cite{polymicro}, 
part number FVP600660710.  These have a 600 $\mu$m diameter ``high $-$OH core'',
a 660 $\mu$m diameter doped silica cladding, and a 710 $\mu$m diameter polyimide outer buffer.  The ends of the
fibers were cut cleanly using a FITEL Model S323 fiber cleaver.

We used an ultraviolet LED of wavelength 270 nm to illuminate individual fibers.  
The bias applied to the LED was in the
range 10.30 V to 11.30 V.  Because of the dynamic range of response we observed over the face of the TPB plate, it
was not possible to use a single bias voltage for the LED for all measurements.  We found that if we controlled the
bias voltage to 1 part in 1000 (that is, to 10 mV), then the output of the LED was reproducible to 1 part in 100.
We were able to establish a relationship between LED bias and light output, and thus we were able to vary the LED bias
as necessary and correct our measurement for the change in light output.

As we only used a single LED for all measurements, it was necessary to be able to position 
one end of each fiber in front of this
LED in a reproducible way.  In fact, the most common ``activity'' during this sequence of measurements was the 
positioning of a fiber in front of the LED.  To make this possible, we found a manufacturer (IDEX) of precision 
stainless steel tubes, where the inner diameter of the tube (760 $\mu$m) 
matched very nicely the outer diameter of the 
fiber (710 $\mu$m).  An individual
fiber was placed in this tube, with one end projecting beyond the end of the tube by about 1 mm, 
and fixed in place with a fast epoxy.  Then a mounting tube was manufactured
and fixed in front of the LED; the inner diameter of the mounting tube was matched to the outer diameter (1/16 inch) of the stainless steel tubes holding the fibers.  
We found, through repeated testing, that we could reproducibly place a fiber in front of the LED and get the same optical throughput, with a 6.5\% uncertainty.

The PMT response displayed a slow, but reproducible, time-dependence (a slow decrease in gain over time), which we observed and were able to correct for.
Quite interestingly, if the high voltage to the PMT was switched off, we observed that the PMT response always returned to the same initial value immediately after the high voltage was switched on again; the time-dependence only affected measurements lasting for more than a few minutes.  It was quite common to switch the high voltage off before changing fibers, so most measurements were taken within one minute of the PMT high voltage having been switched on.  We did not discover the underlying cause of this time-dependence; we can however say that it is not due to a changing temperature of the PMT or base, because they were submerged in liquid argon.

\begin{figure}[ht]
\begin{center}
\includegraphics[scale=0.35]{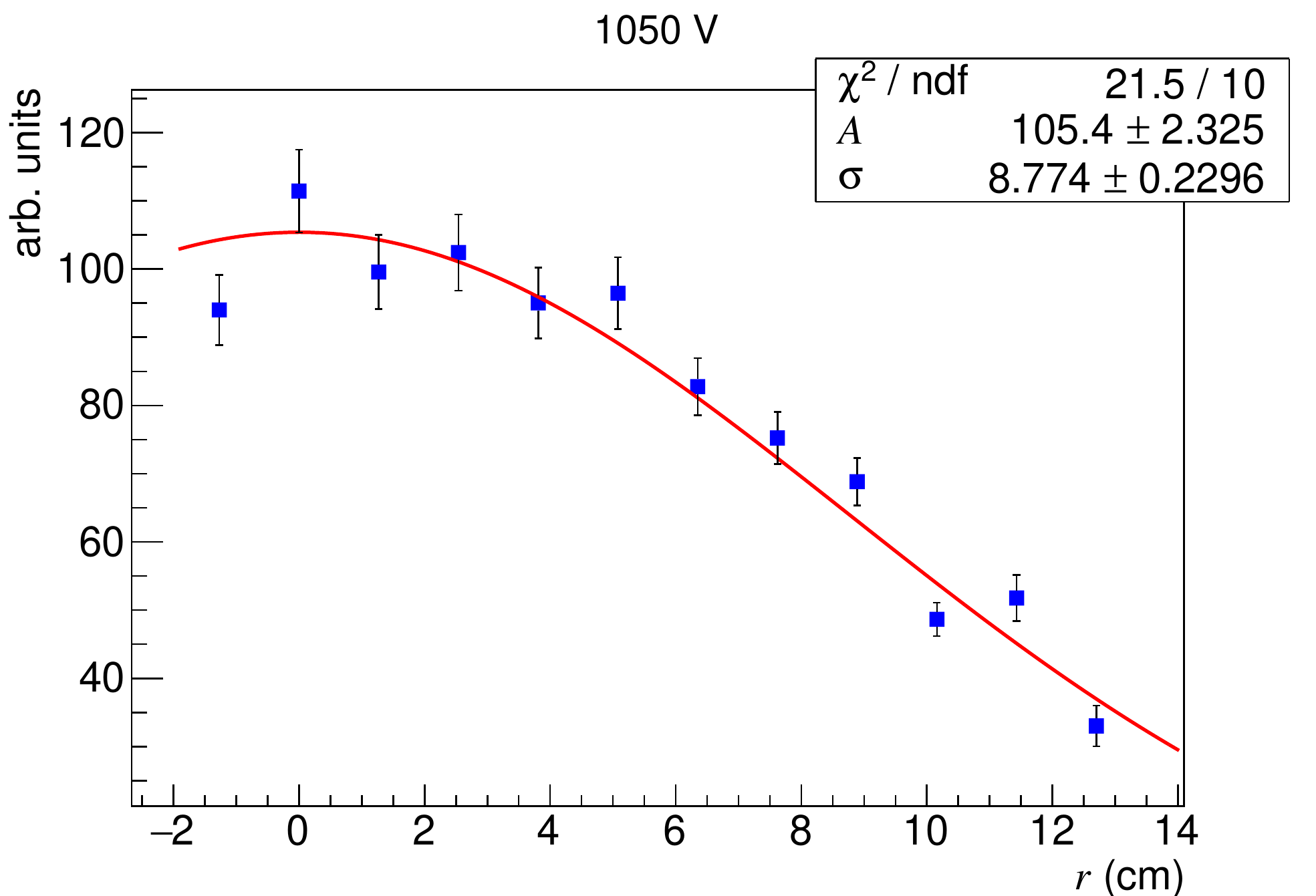} \includegraphics[scale=0.35]{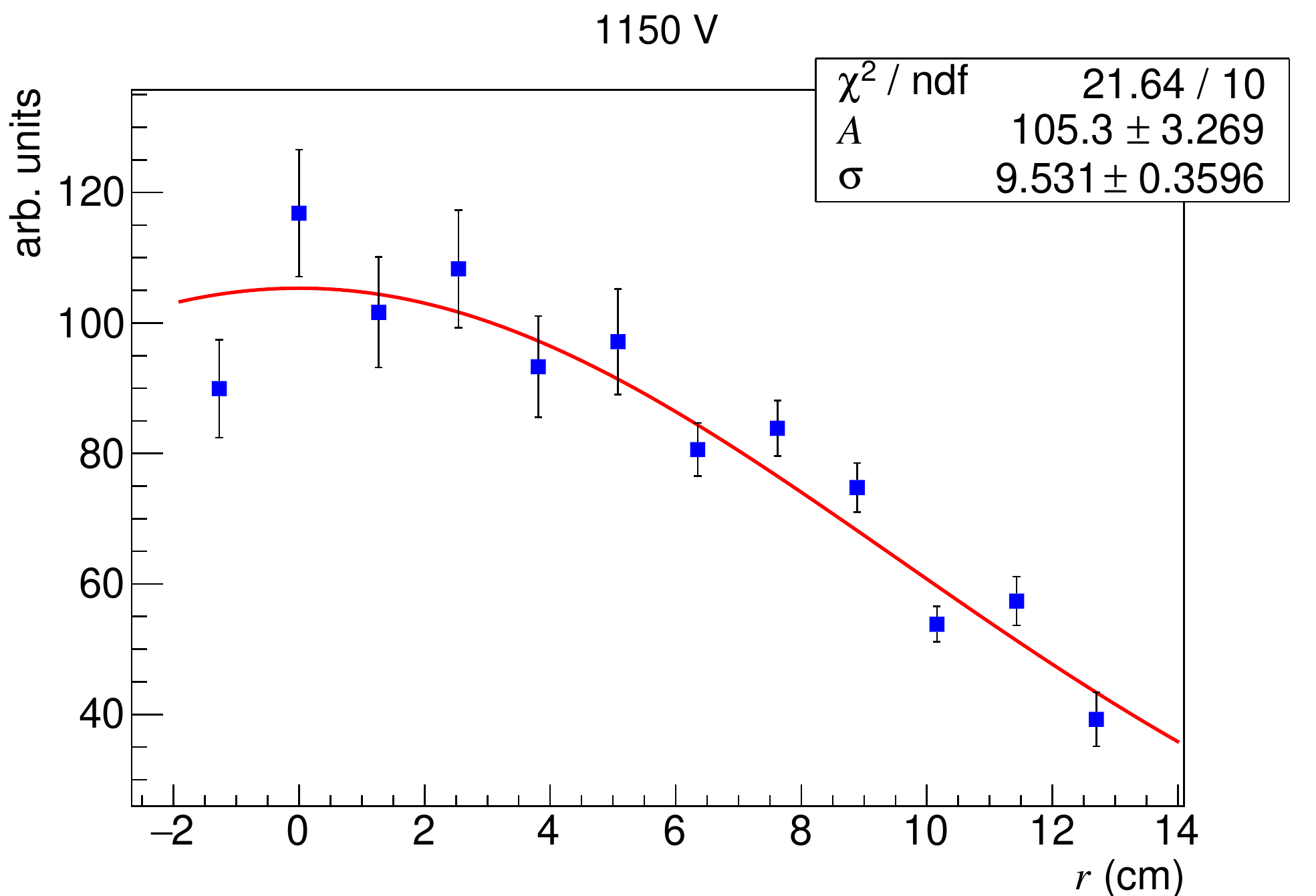} \\
\includegraphics[scale=0.35]{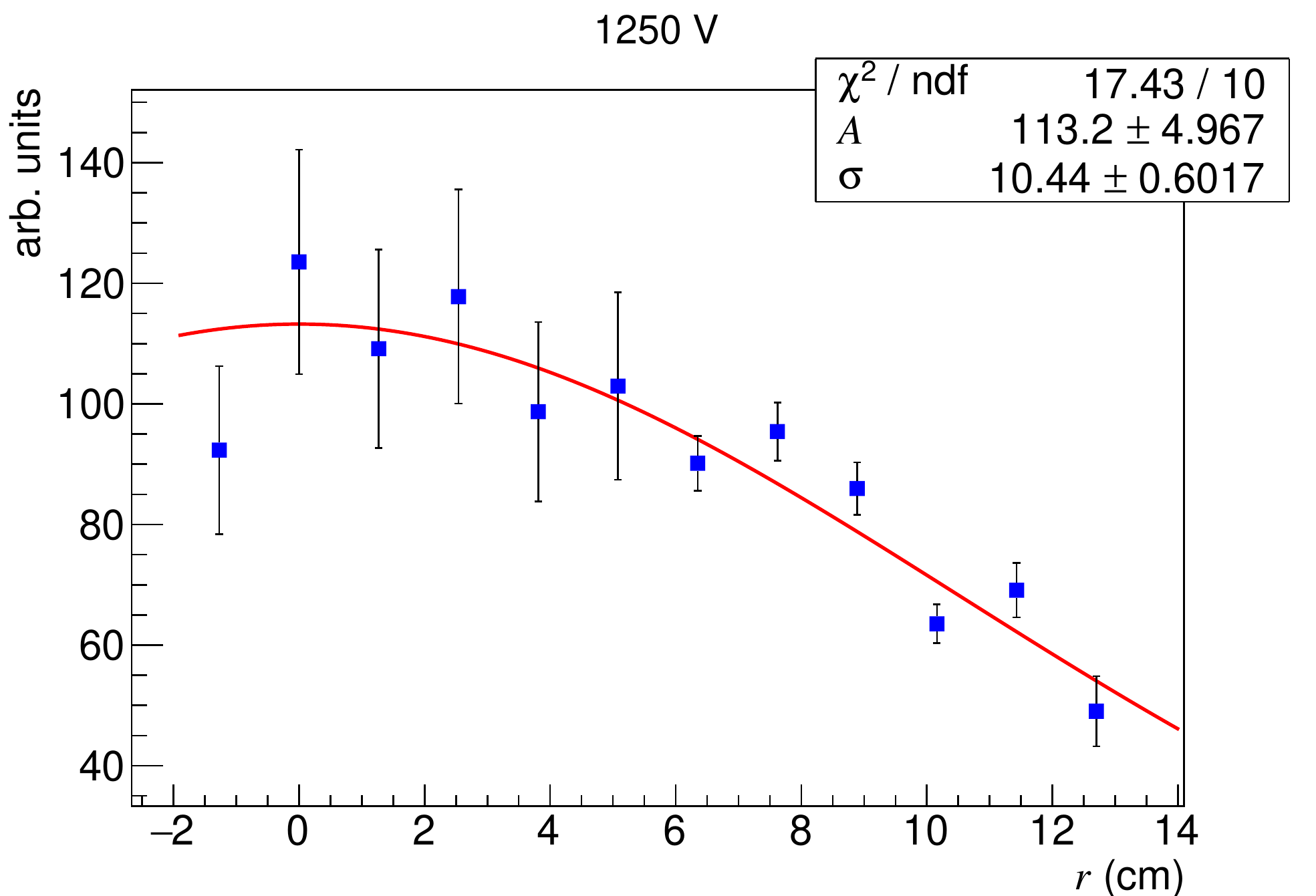} \includegraphics[scale=0.35]{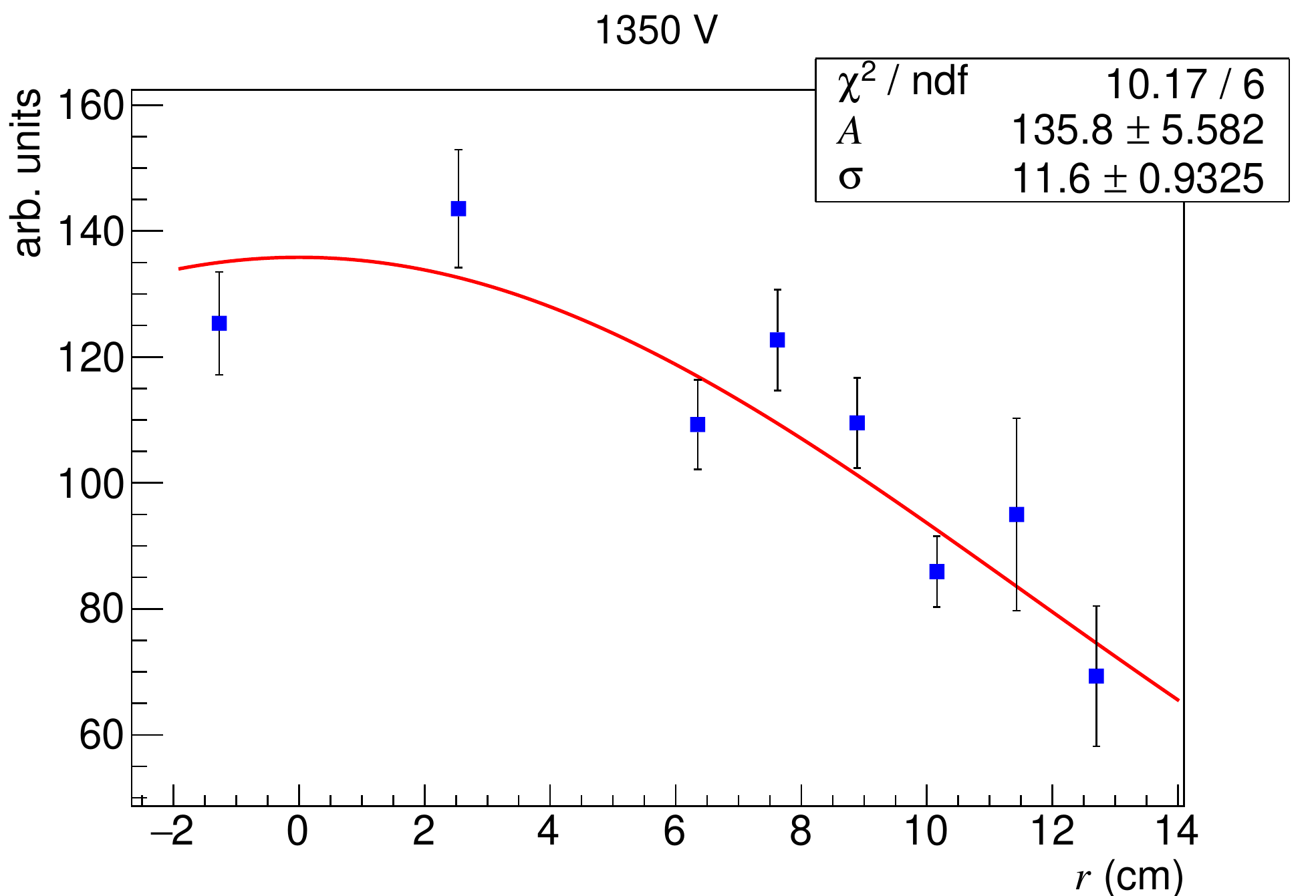}
\end{center}
\caption{Response of the optical unit as a function of radius (cm) from the center of
the TPB plate, for four different high voltages applied to the PMT.  The uncertainties are from a variety
of corrections, described in the text.  The red curve is a fit to a Gaussian function, 
$f(r) = A\exp[-r^2/2\sigma^2]$.  It
is seen that the width, $\sigma$, increases from about 8.8 cm to 11.6 cm as the PMT high voltage
is increased from 1050 V to 1350 V.}
\label{results_figure}
\end{figure}

We observed the PMT response at four values of the PMT high voltage: 1050, 1150, 1250, and 1350 V.  At each
of these voltages, we illuminated each of the 12 fibers using the UV LED and recorded the average anode current pulse 
height.  The LED was pulsed at a rate of 100 Hz.  Cosmic rays passing through the liquid argon also produced light 
pulses, but the resulting signals were smaller than those produced by the LED flashes and did not represent 
a significant background to this measurement.  The results of this study are shown in figure~\ref{results_figure}.
The statistical contribution to the uncertainties is insignificant; the uncertainties displayed are 
purely systematic and arise from four sources: [1] corrections due
to the changes in LED bias voltage (1\% uncertainty); [2] corrections due to small gain shifts between different measurement periods ($\approx2$\% uncertainty);
[3] corrections due to the measured fiber-to-fiber differences ($\approx10$\% uncertainty); and
[4] uncertainty in transmission when changing from
one fiber to the next (6.5\% uncertainty).
It is seen that the response varies by a factor of 3
over a distance of 13 cm in radius.  A fit to each response function, using a Gaussian shape, was performed,
and the results may be seen in figure~\ref{results_figure}.  There is a trend towards a flatter response
function with increasing PMT high voltage; this could be attributable to a change in photoelectron collection
optics.

To study azimuthal variations in the response, we held the PMT at a fixed value of high voltage, illuminated a single
fiber, and rotated the optical unit in a series of steps 
for a total of 8 measurement points.  
We did this using two different fibers:
\begin{itemize}
\item A fiber located 2.54 cm from
the center of the TPB plate; 8 points were illuminated, spaced 12$^\circ$ apart.
\item Another fiber located 10.2 cm from the center; 8 points were illuminated, spaced 24$^\circ$ apart.
\end{itemize}
(We were not able to rotate the PMT more than 170$^\circ$ due to
constraints imposed by interferences between cables and support structures.)
We observed variations of not more than 20\% from one point to another, in these two azimuthal scans.    
Significantly, these variations are not larger than those arising from the PMT itself as seen in the
studies performed in references \cite{Koblesky201240} and \cite{Abbasi2010139}.  We see no evidence
for significant variation in the surface quality of the TPB coating.  We set a limit of $|\delta| < 0.1$ for the
azimuthal variation parameter.

\section{GEANT4 Simulation}\label{sec:sim}

We created an optical photon simulation of the experiment described in Section \ref{sec:absnorm} using the software package GEANT4~\cite{GEANT}. The simulation incorporates the geometry of the experiment, and the relevant physics processes for photon ray tracing. We divide the simulation into three distinct steps, each corresponding to one physical process:

\begin{enumerate}
	\item Alphas are emitted from the surface of the polonium source disk.
	\item Scintillation photons from alphas travel through the liquid argon volume and may hit the TPB plate.
	\item Photons which hit the TPB plate have a probability of re-emission, and re-emitted photons may hit the PMT and be converted to photoelectrons.
\end{enumerate}

\subsection{Simulation Description}

\textit{Step 1}: Due to the geometry of the cryostat and the source holder (discussed in \cite{Jones-JINST-8-07-P07011}), the number of alpha scintillation photons which hit the TPB plate depends on the position of each alpha emission on the source disk. To simulate alpha emission from the surface of the source disk, alpha emission positions are chosen from a Gaussian distribution centered on the plane of the source disk. At these positions, we put an isotropic point source of 128 nm light into the simulation.

\begin{figure}
	\begin{center}
	\begin{subfigure}{0.32\textwidth}
		\begin{minipage}{1.0\textwidth}
			\includegraphics[width=1.0\textwidth]{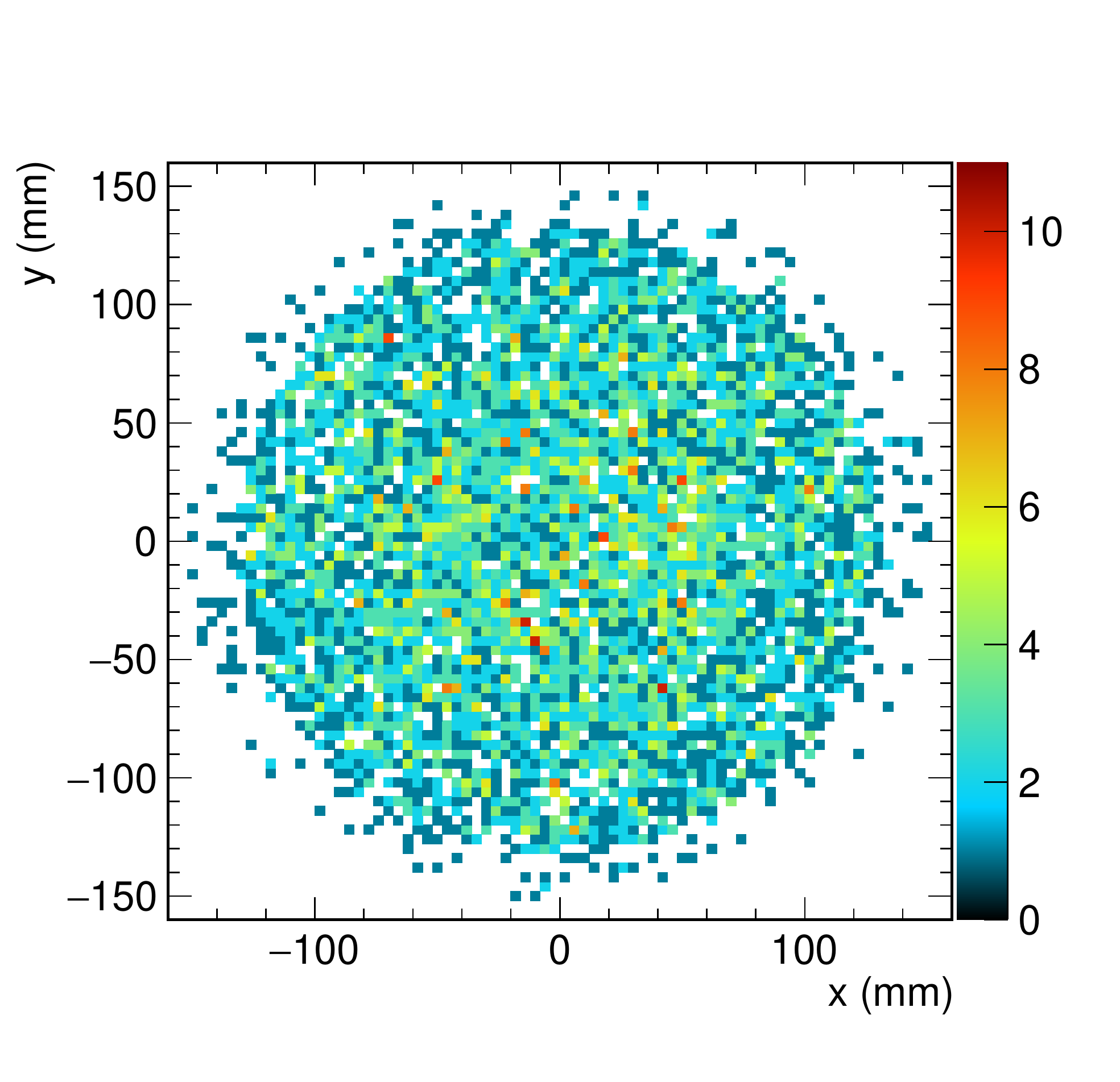}
            \includegraphics[width=1.0\textwidth]{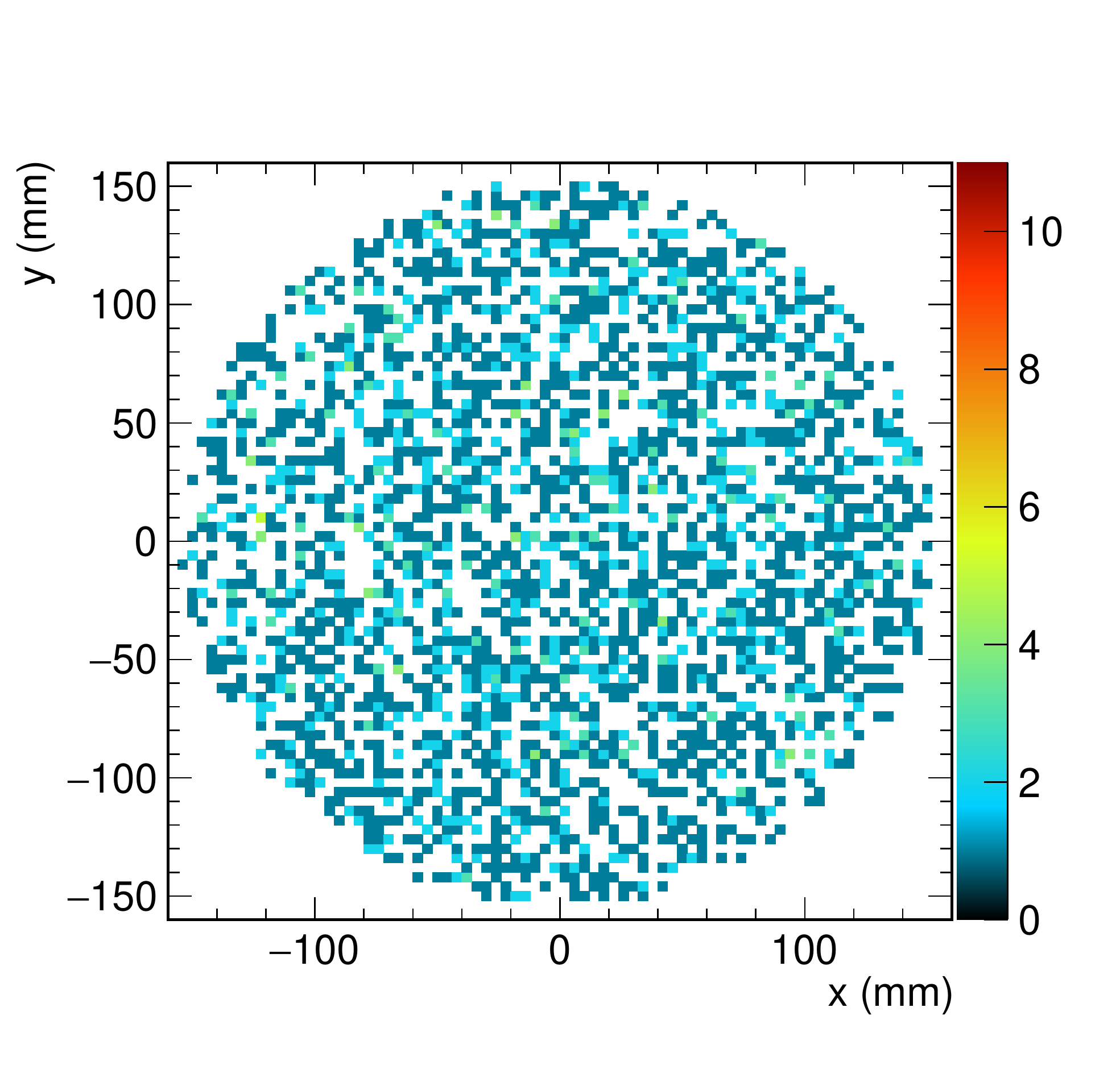}
		\end{minipage}
		\caption{$r=0$ mm}
	\end{subfigure}
	\begin{subfigure}{0.32\textwidth}
		\begin{minipage}{1.0\textwidth}
			\includegraphics[width=1.0\textwidth]{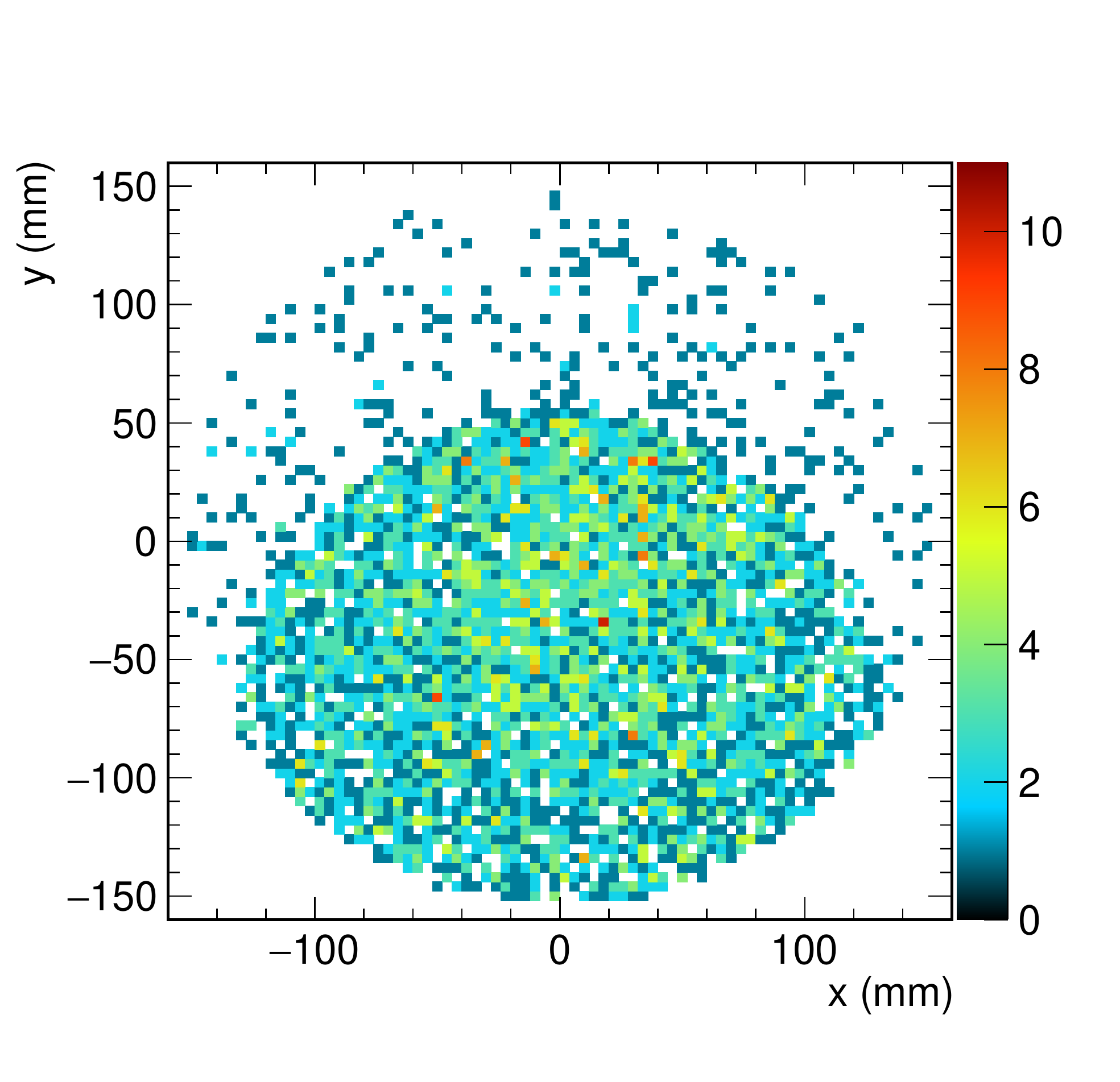}
            \includegraphics[width=1.0\textwidth]{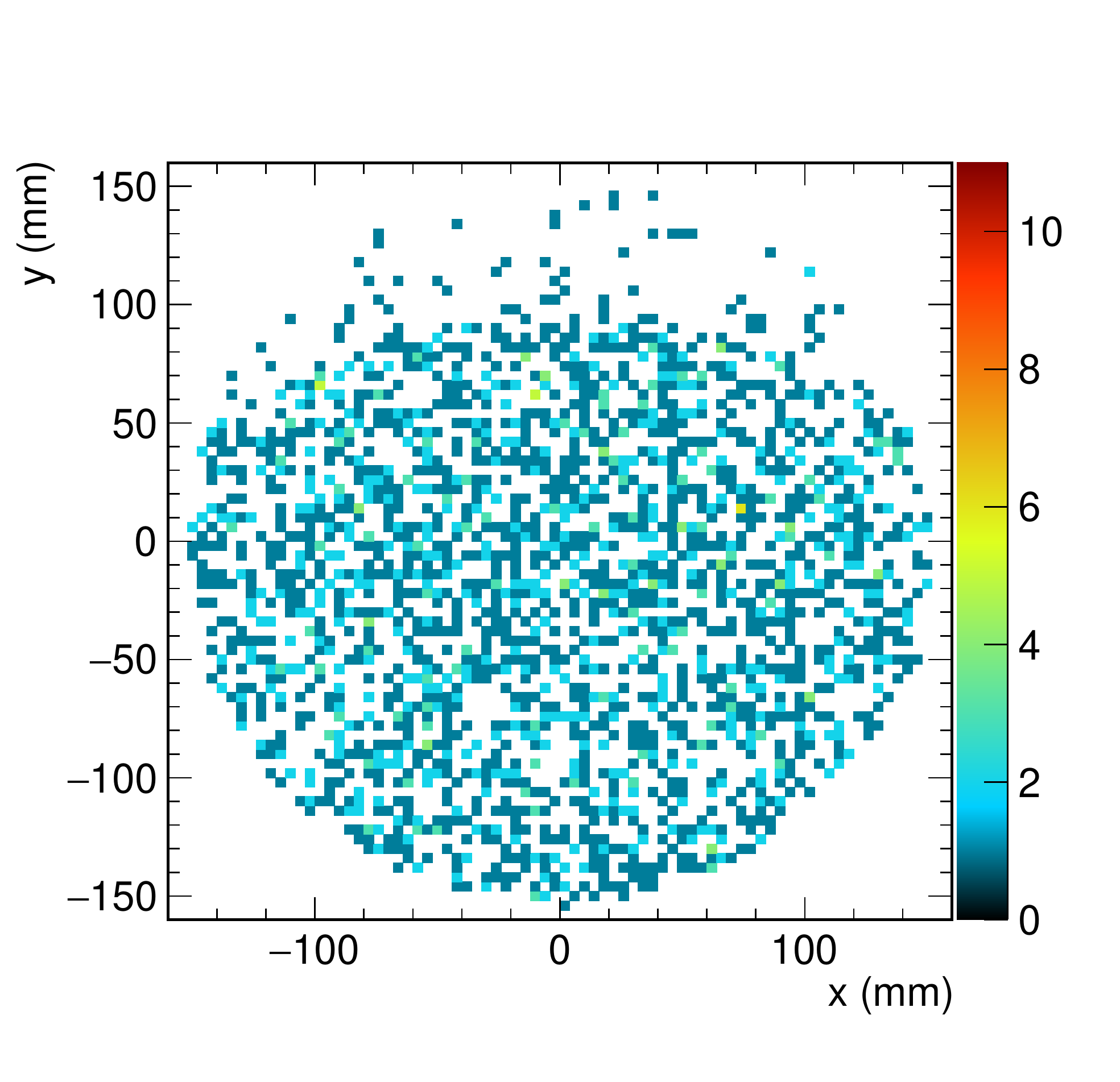}
		\end{minipage}
		\caption{$r=1.5$ mm}
	\end{subfigure}
	\begin{subfigure}{0.32\textwidth}
		\begin{minipage}{1.0\textwidth}
			\includegraphics[width=1.0\textwidth]{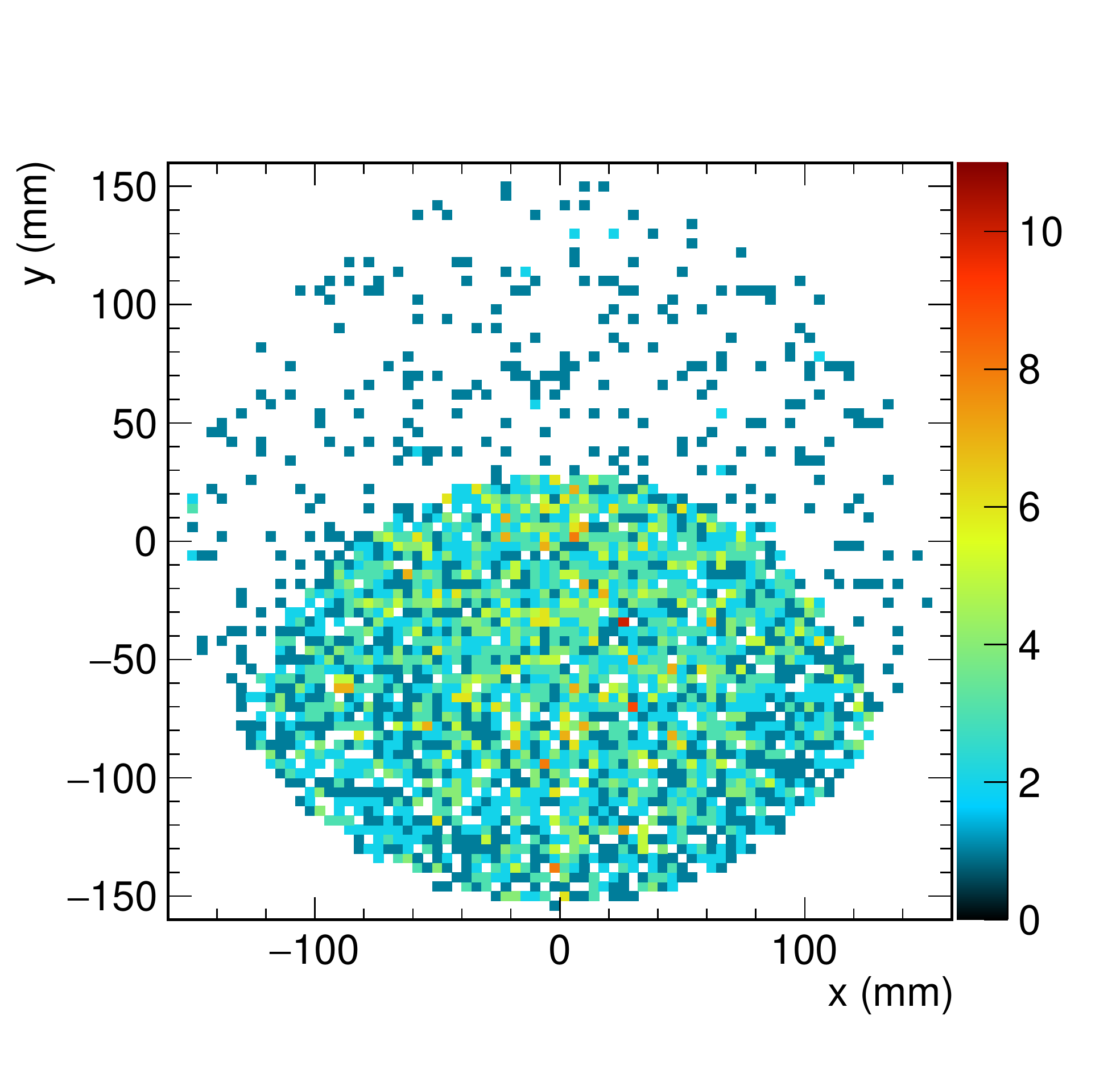}
            \includegraphics[width=1.0\textwidth]{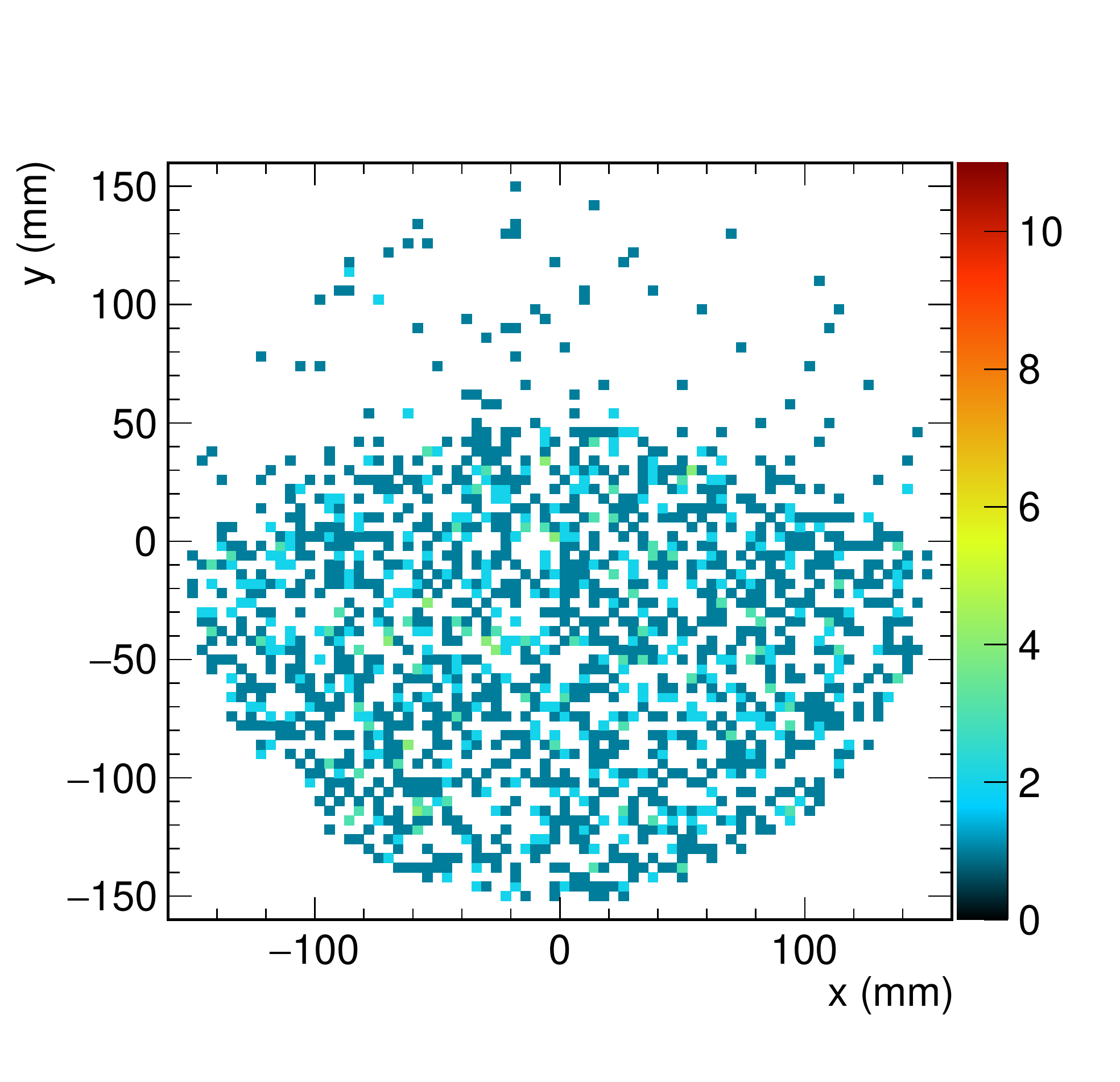}
		\end{minipage}
		\caption{$r=2.0$ mm}
	\end{subfigure}
	\end{center}
	\caption{Simulated photon hit positions (units mm) on a TPB plate for a source placed at $D=20.3$ cm (top row) and $D=36.8$ cm (bottom row) as a function of the alpha emission position on the source disk, $r$. Shadowing on the TPB plate is observed as the emission point is moved behind the collimator of the source holder.\label{fig:hitpos}} 
\end{figure}

\vspace{0.5em}

\begin{figure}
\begin{center}
	\begin{subfigure}{0.49\textwidth}
		\begin{minipage}{1.1\textwidth}
			\includegraphics[width=1.0\textwidth]{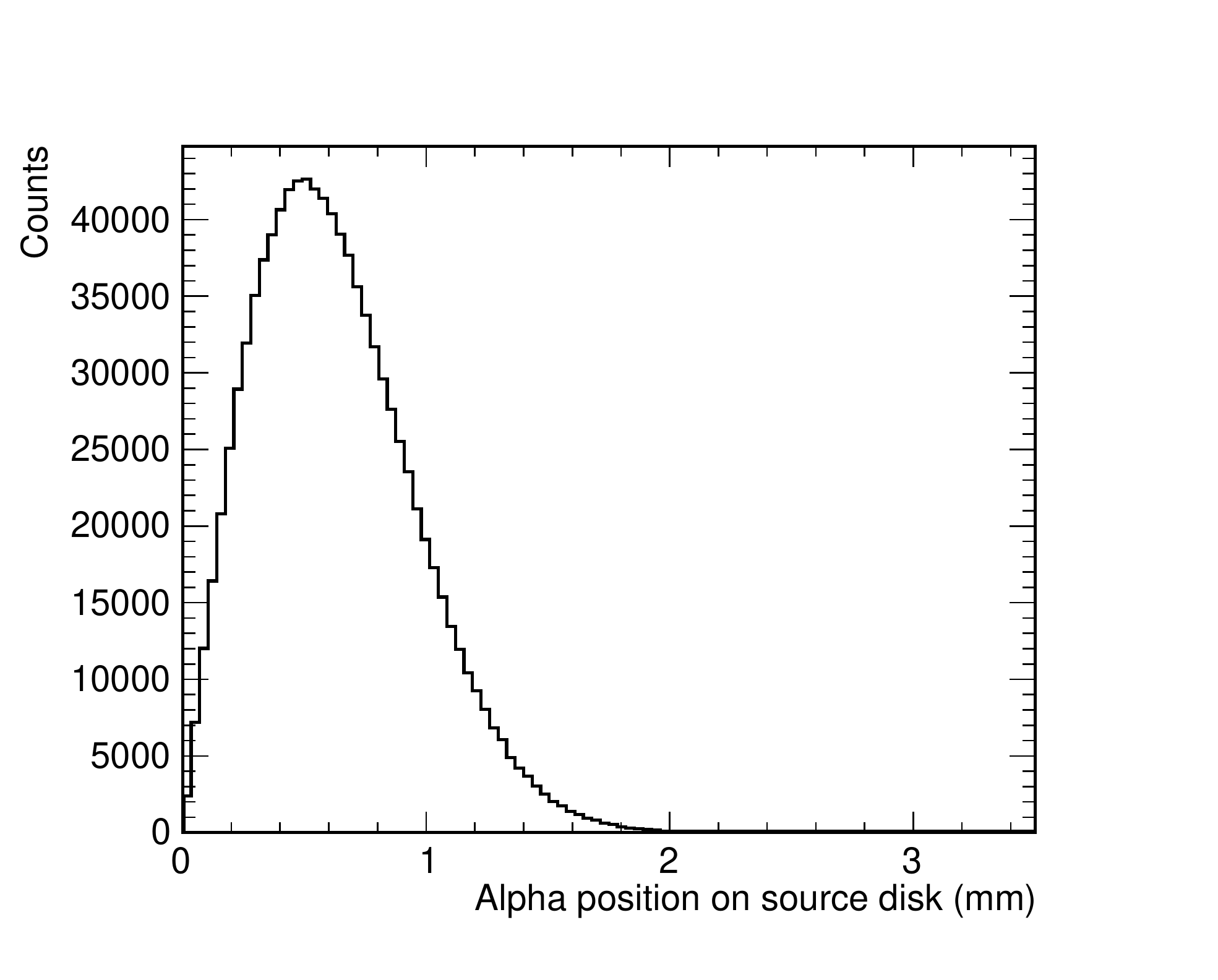}
		\end{minipage}
		\caption{Alpha position distribution from the simulation.}
	\end{subfigure}
	\begin{subfigure}{0.49\textwidth}
		\begin{minipage}{1.1\textwidth}
			\includegraphics[width=1.0\textwidth]{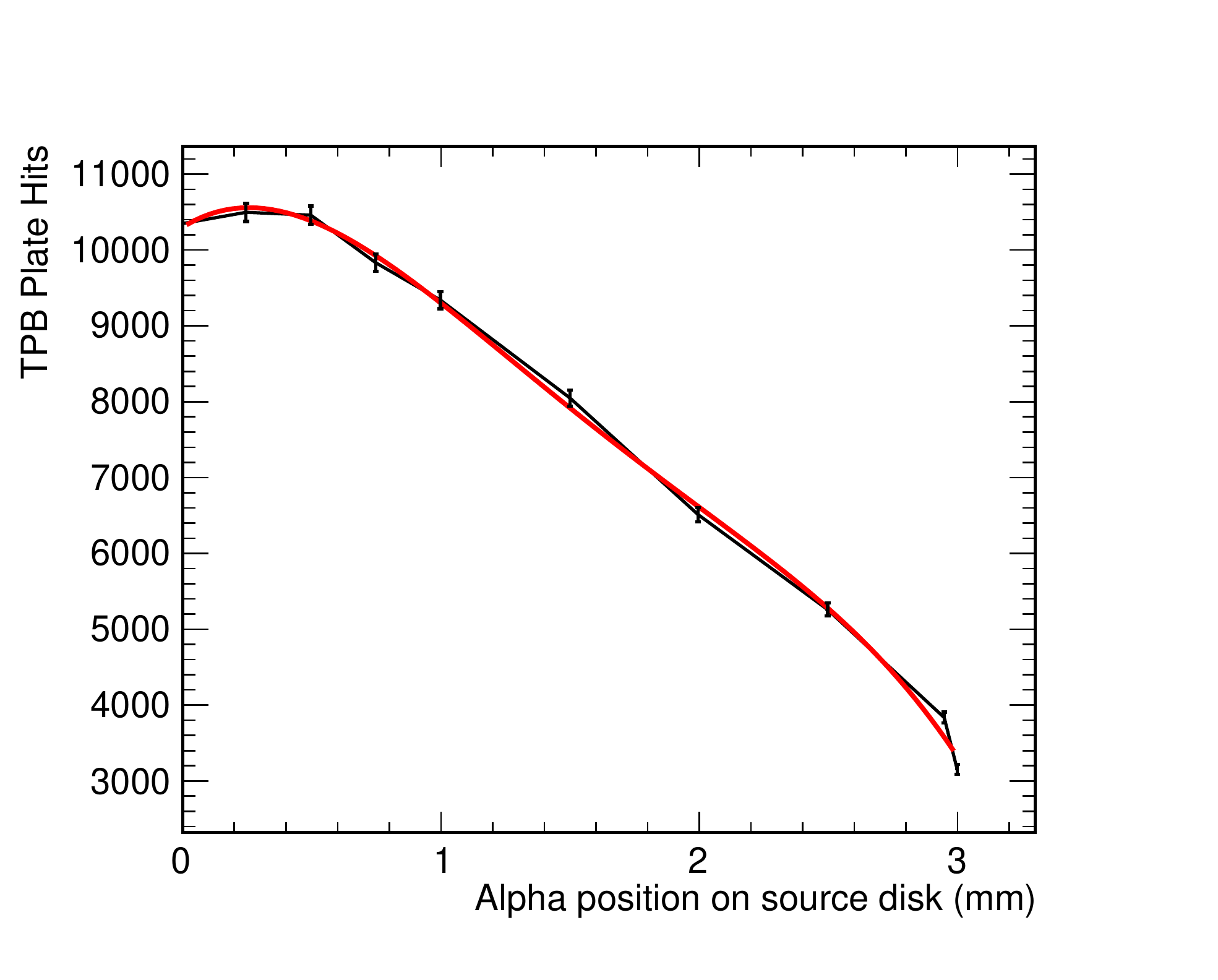}
		\end{minipage}
		\caption{Fitted TPB plate hits vs. alpha position}
	\end{subfigure} 
    \begin{subfigure}{0.7\textwidth}
		\begin{minipage}{1.0\textwidth}
			\includegraphics[width=1.0\textwidth]{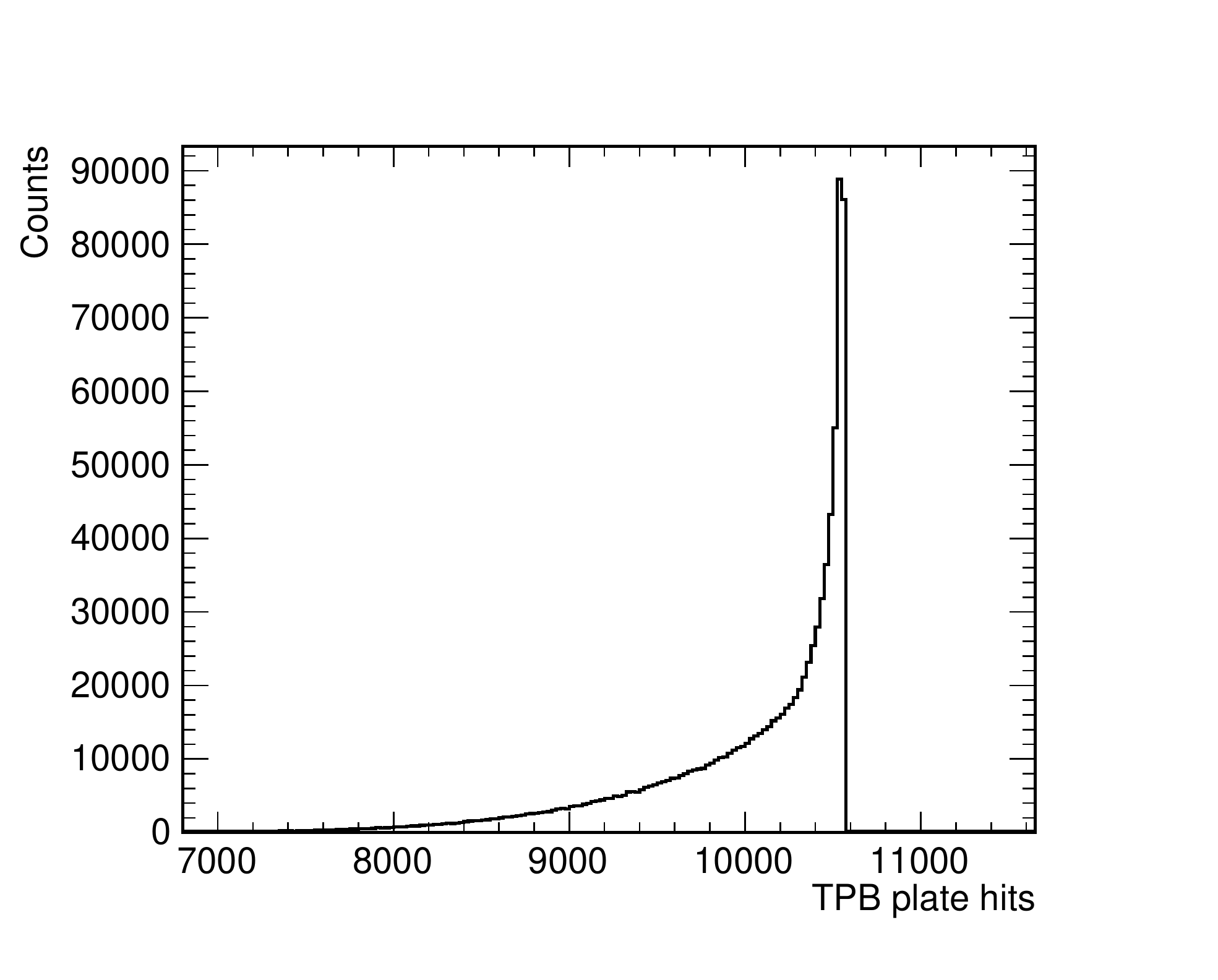}
		\end{minipage}
		\caption{Convolution result: TPB plate hit spectrum}
	\end{subfigure} 
\end{center}
  \caption{(a): Simulated radial distribution of alpha emissions. The width of this distribution is controlled by a simulation parameter, listed in table~\ref{tab:params}. (b): Fitted (red) TPB plate hits vs. alpha emission position curve. The black points in this plot are from a fixed set of alpha emission positions (the variable $r$ in figure \ref{fig:hitpos}) chosen to efficiently map the trend in hits; it is infeasible to do simulations for each possible alpha position. (c): Convolution result of (a) and (b), the spectrum of TPB plate hits. The spectrum is tightly peaked near 10600 hits for the parameters chosen; the tail to lower numbers is due to the population of alphas emitted near the edge of the source disk, which are shadowed by the source collimator.}
  \label{fig:tpbhits}
\end{figure}

\noindent\textit{Step 2}: We track the photons produced at each alpha position to the TPB plate, and record the number and positions of photons which hit the plate. Several examples of these plate hit position distributions for various alpha positions are shown in figure \ref{fig:hitpos}. Due to computing limitations, we approximate the dependence of the number of plate hits on alpha position by choosing a subset of alpha starting positions along the radius of the source disk, and simulating the TPB plate hits from alpha emissions only at those positions. We then fit a smooth curve to the number of hits at each alpha starting position. The fitted curve is convolved with the distribution of alpha positions from step 1 to obtain the spectrum of plate hits. The radial distribution from the simulation, the fitted plate hit curve, and the convolution result are shown in figure \ref{fig:tpbhits}(a-c). We scale the plate hit curve (figure \ref{fig:tpbhits}b) by the expected number of scintillation photons per alpha, to correct for the number of photons generated in the simulation. To obtain the expected number of plate hits from the simulation, we take the mean of the plate hit spectrum, shown in figure \ref{fig:tpbhits}c.

\vspace{0.5em}

\begin{figure}
  \begin{center}
  	\includegraphics[width=0.8\textwidth]{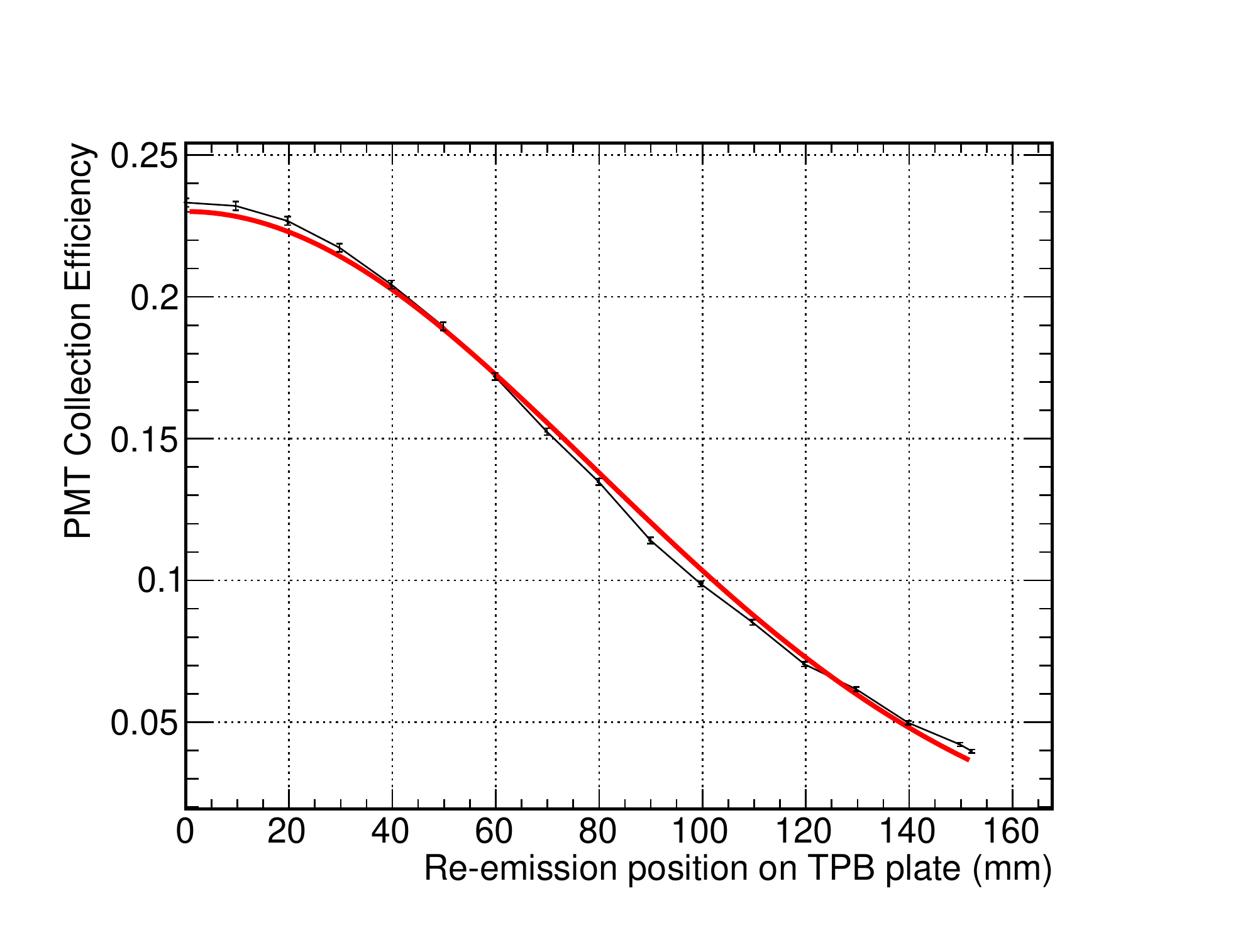}
  \end{center}
  \caption{Simulated PMT collection efficiency (number of optical photons originating in the plate which reach the PMT, divided by the total number of photons) as a function of re-emitted photon position along the TPB plate radius. The data points are fitted to a mean-0 Gaussian function (red) in order to compare the width to the experimental curves shown in figure \ref{results_figure}. The width from the fit shown is $7.9 \pm 0.2$ cm, and is nearly consistent with the 1050-V measurement from figure \ref{results_figure}.}
  \label{fig:plateeff}
\end{figure}

\noindent\textit{Step 3}: We use the plate hit position distributions from the previous step to simulate a PE spectrum. As discussed in section \ref{sec:posdep}, the probability of a re-emitted photon hitting the PMT depends on the location of re-emission in the TPB layer of the plate. Similarly to how a fixed set of alpha positions was chosen in the previous simulation step, we choose a fixed set of points along the radius of the TPB plate, and generate populations of isotropic 420 nm photons in the TPB layer of the acrylic plate at these positions. These photons are ray-traced through the plate, and we record the ratio of photons which reach the PMT surface to the total number of photons generated. In this way, we generate a curve which describes the optical photon collection efficiency from the TPB plate, as a function of re-emission position on the plate. This efficiency curve is shown in figure \ref{fig:plateeff}. We note the strong similarity of the shape of the simulation results to the experimental curves shown in figure \ref{results_figure}; this is a very important indicator that our simulation correctly describes the distribution of UV photons hitting the TPB plate and the subsequent production of optical photons.

To simulate the PE spectrum, we take the recorded TPB hit position distributions, generated in step 2, for each alpha starting position, and convolve the distributions with the TPB plate efficiency curve. This gives the expected number of PMT hits as a function of alpha position on the source disk. We then fit this dependence to a smooth curve, and scale it by several conversion factors to convert PMT hits into PE. The curve is shown in figure \ref{fig:pevsr}. The conversion factors we consider are: 

\begin{figure}
  \begin{center}
  	\includegraphics[width=0.8\textwidth]{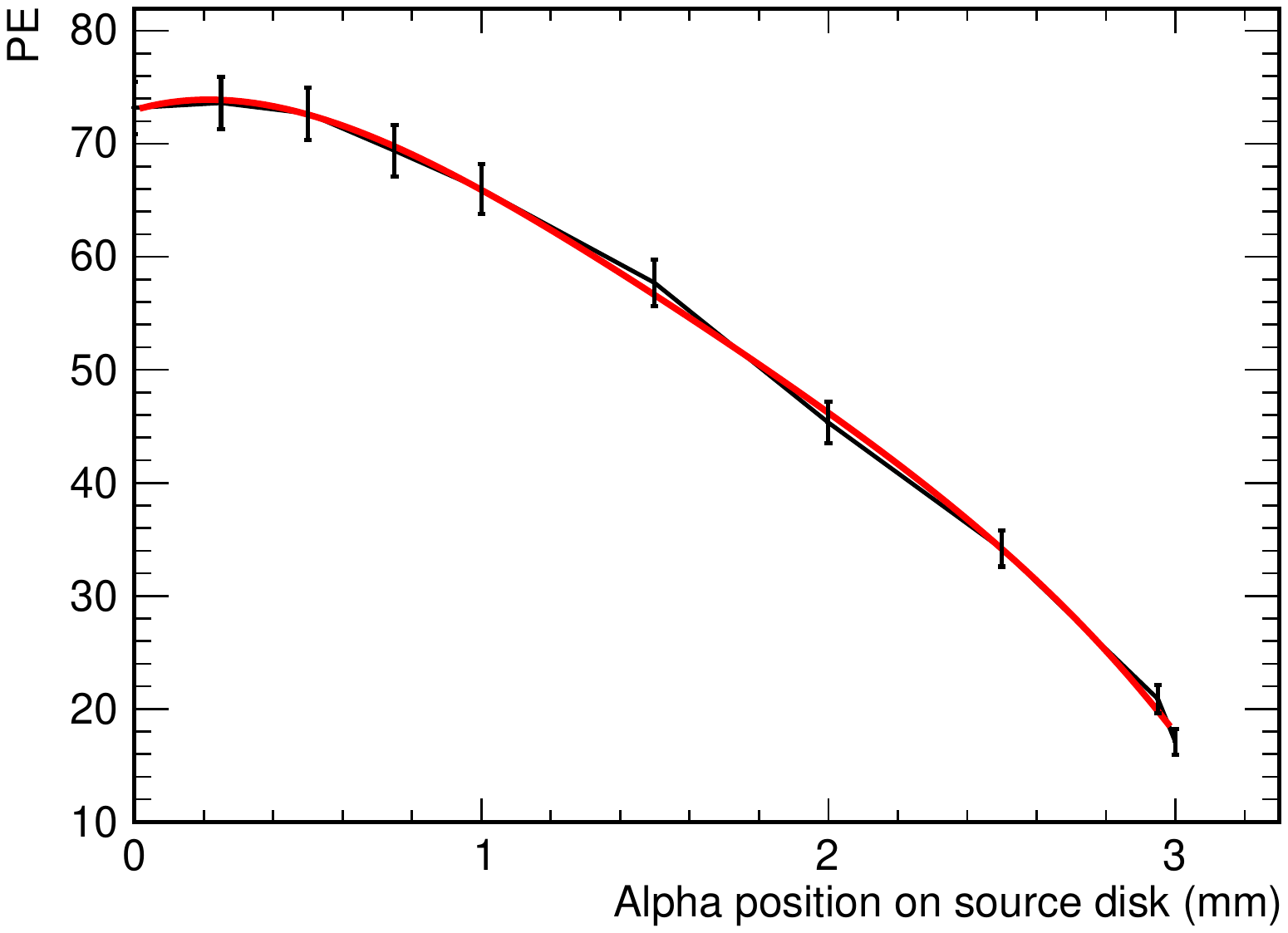}
  \end{center}
  \caption{Dependence of the mean number of photoelectrons produced on the position of alpha starting position in the source holder. The points are fitted to a smoothing function (red). This function is convolved with the distribution of alpha positions from the simulation to produce the spectra in figure \ref{fig:pespec}.}
  \label{fig:pevsr}
\end{figure}

\begin{itemize}
	\item The number of scintillation photons per alpha emission;
	\item The probability of re-emission for a plate coated via evaporation of TPB at room temperature;
	\item The relative probability of re-emission for a plate dipped in TPB to that of one coated via evaporation;
    \item Scaling factor to convert the re-emission probability measured at 290 K to 87 K;
	\item The quantum efficiency of the PMT at 87 K.
\end{itemize}

\begin{figure}
\begin{center}
	\begin{subfigure}{0.49\textwidth}
		\begin{minipage}{1.1\textwidth}
			\includegraphics[width=1.0\textwidth]{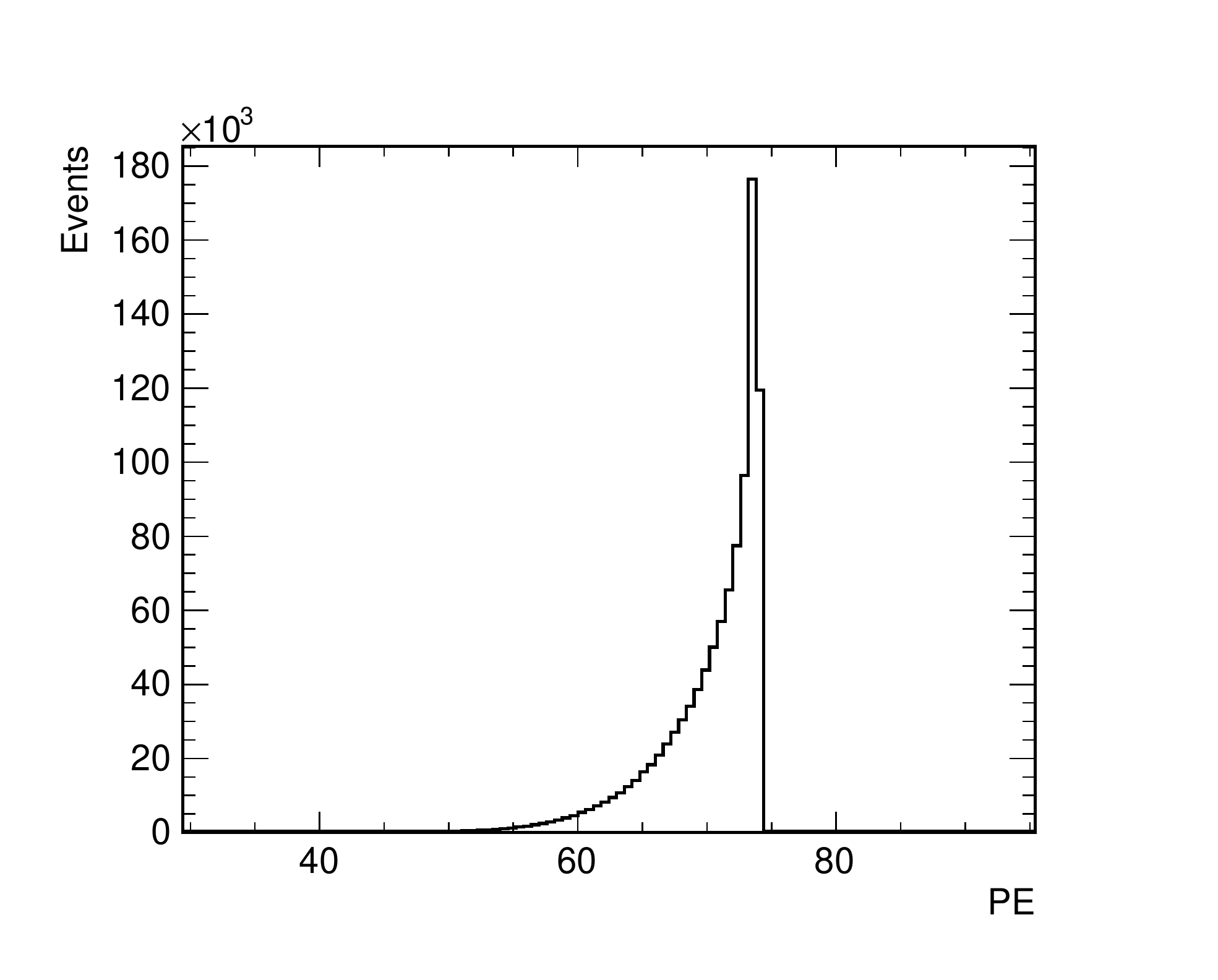}
		\end{minipage}
		\caption{Simulated PE Mean Spectrum}
	\end{subfigure}
	\begin{subfigure}{0.49\textwidth}
		\begin{minipage}{1.1\textwidth}
			\includegraphics[width=1.0\textwidth]{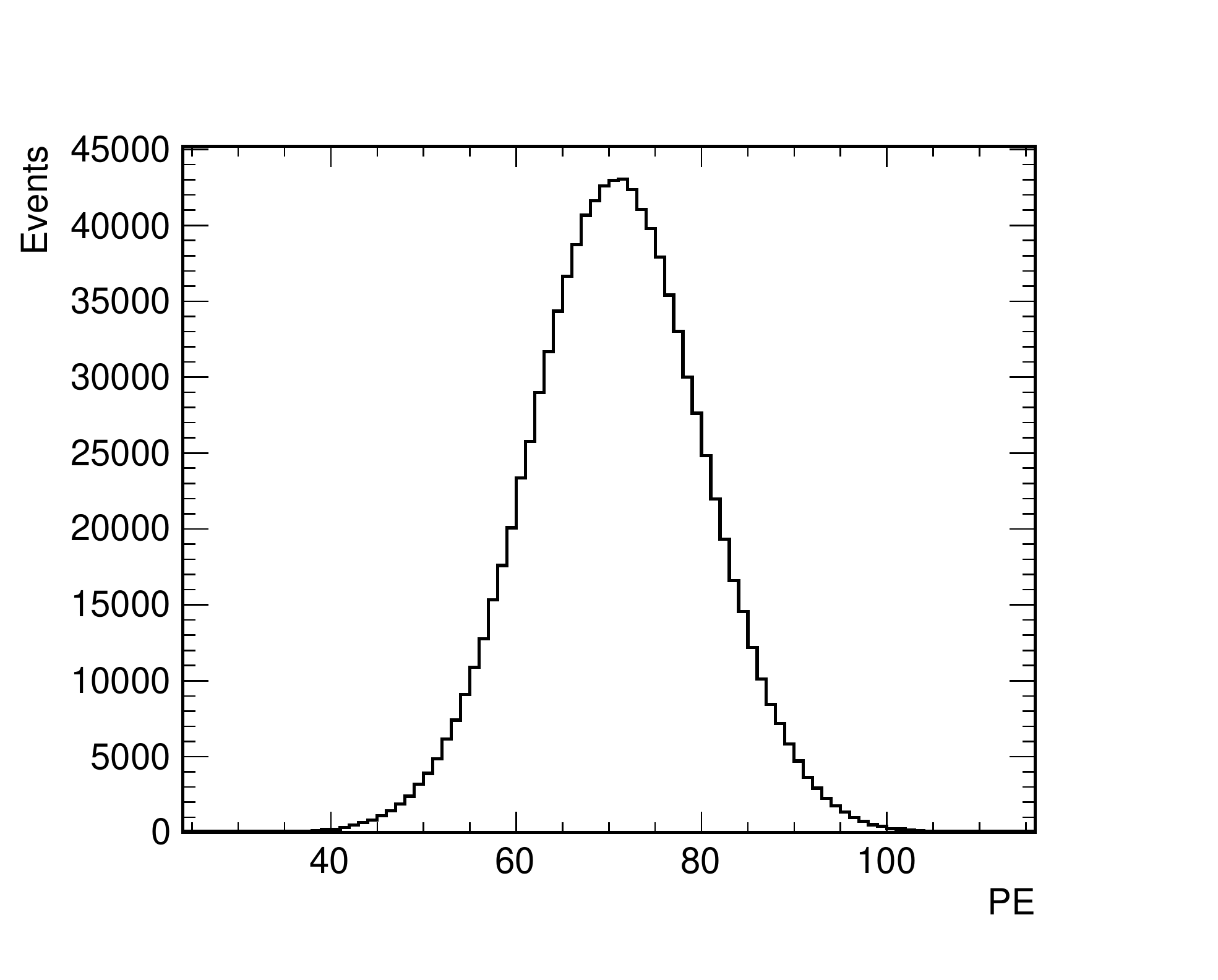}
		\end{minipage}
		\caption{Poisson-Smeared Mean Spectrum}
	\end{subfigure} 
\end{center}
  \caption{(a): One example of a simulated PE Mean spectrum, resulting from the convolution of the distributions of alpha positions with the function from figure \ref{fig:pevsr}. (b): The associated Poisson-smeared spectrum, to simulate observed data. The peak of this spectrum is fitted to determine a simulated value for $N_{\rm PE}$, which is then compared to data.}
  \label{fig:pespec}
\end{figure}

Finally, we convolve this scaled, fitted curve with the distribution of alpha positions from step 1. The resulting histogram is the simulated mean-PE spectrum. Because this spectrum is a mean spectrum, we simulate a Poisson process for each PE we fill in the histogram. This gives the simulated PE spectrum which we can compare to the data. The resulting PE spectra from this step is shown in Figure \ref{fig:pespec}. We fit the Poisson-smeared PE spectrum with a Gaussian around the peak to obtain the peak position, which is the simulated value of $N_{\rm PE}$. We will compare this number to the measurements from Section \ref{sec:absnorm}.

\subsection{Simulation Uncertainties}

Because the simulation results depend on a number of geometric and optical properties, we need a robust method for determining the associated uncertainty and central value for the $N_{\rm PE}$. We assume Gaussian prior probability distributions for each parameter included in the simulation, and then re-run the full simulation multiple times, using new parameters drawn from these prior distributions. This demonstrates the spread in simulated $N_{\rm PE}$ values. In total, 1300 unique parameter sets were tested with the source at the near position, and 900 sets were tested with the source at the far position.

The central values for the parameters which have the most influence on the simulation result, along with each parameter's 1-$\sigma$ value, are listed in Table \ref{tab:params}. An explanation of some of these parameters is given below:

\begin{table}[t]
	\begin{center}
		\begin{tabular}{ ccc }
			\textbf{Parameter} & \textbf{Central Value} & \textbf{1-$\sigma$}\\
			\hline
			$R_{\text{Al}}$, 128 nm \cite{bricola_2007}	&	0.12	&	$\pm 0.02$\\
			$R_{\text{Stainless}}$, 128 nm \cite{karlsson_ribbing_1982} &	0.35	&	$\pm 0.05$	\\
			$N_{\text{LAr}}$, 128 nm \cite{grace_butcher_monroe_nikkel_2017}	&	1.45	&	$\pm 0.07$\\
			$N_{\text{LAr}}$, 420 nm \cite{sinnock_smith_1969}	&	1.23	&	$\pm 0.002$\\
			$N_{\text{Acrylic}}$, 420 nm \cite{sinnock_smith_1969} &	1.49	&	$\pm 0.02$\\
			$N_{\text{Glass}}$, 420 nm \cite{malitson_1965}	&	1.46	&	$\pm 0.04$\\
            Rayleigh Parameter, $\lambda$ \cite{grace_butcher_monroe_nikkel_2017}	&	60 cm	&	$\pm 6$ cm\\
			\hline
			Source Distribution Width	&	0.08 mm	&	$\pm 0.07$ mm  \\
			Plate to PMT Distance	&	0.125 in	&	$\pm 0.03$ in\\
			Glass Thickness	&	5 mm	&	$\pm 1$ mm\\
			\hline
			Scint. photons per Alpha	&	134000	&	$\pm 6000$\\
            Evap. TPB Efficiency \@ 290 K\cite{benson_2017} &	0.40	&	$\pm 0.04$\\
            Temperature Correction \cite{francini_2013} & 1.22 & $\pm 0.15$ \\
			Rel. Efficiency to Evap. TPB, 87 K \cite{ignarra:2014} &	0.67	& $\pm 0.06$		\\
			PMT Quantum Efficiency, 87 K \cite{Acciarri:2016smi}	&	0.153	&	$\pm0.008$	\\
		\end{tabular}
		\caption{Parameters, with associated uncertainties, considered for the GEANT4 simulation.}
        \label{tab:params}
	\end{center}
\end{table}

\begin{itemize}
	\item \textit{Source Distribution Width}: The simulation assumes the polonium has a Gaussian distribution over the source disk, such that alphas are more likely to be emitted close to the center than at the extreme edges. We choose this value to simulate a small amount of ``shadowing'' of the TPB plate caused by the source holder in the experimental setup. We note that, despite the large physical implication of this parameter, the fitted PE spectrum peak does not depend strongly on this. Larger values of this parameter may move the peak lower due to the associated increase in size of the low-PE tail, but the effect is small. Furthermore, as per Table \ref{tab:params}, we allow a large variance in this value, so its effect is captured when we sample many parameter sets.
    
	\item \textit{$R_{\rm Al}$, 128 nm}: In \cite{bricola_2007}, the authors list 14\% for 172 nm light, and we have a lower limit from \cite{johnson_1968} of 10\%. We choose the average value of these two numbers, with the understanding that 14\% is conservative.

	\item \textit{Scintillation Photons per Alpha}: Liquid argon has a maximum scintillation yield of $51000\pm 1000$ scintillation photons per MeV of energy deposited~\cite{DOKE1988291}.  The scintillation from alpha particles is significantly quenched; the yield from alpha particles is $71\pm 2$\% of the maximum~\cite{DOKE1988291}. Additionally, not all of the light shows up in the prompt component; Lippincott et al.~\cite{Lippincott-PhysRevC.78.035801} measured the prompt light component to be $70 \pm 2$\% of the total.  Therefore we expect a total of $N_\gamma = 134000 \pm 6000$ prompt scintillation photons per alpha particle.

	\item \textit{TPB Efficiencies}: The measurement in \cite{Gehman2011116} was recently revised by a new experiment described in \cite{benson_2017} and was changed from $1.18$ to $0.40$. While testing the simulation, we were unable to reproduce the correct PE peak position using the old value. When using the updated value, our simulated peak positions tended to fall slightly below the PE peak measured, but were significantly closer, and were well within the range of the measurements.
    
    \item \textit{Relative Efficiency to Evaporated TPB, at 87 K:} We note that the TPB efficiency measurement was made at room temperature and with the correct thickness of our TPB coating sample (1.8 $\mu$m). However, the method for coating the plates was not the same -- the plates tested were coated evaporatively, while the MicroBooNE plates in our experiments were coated via dipping. A measurement~\cite{ignarra:2014} was made to compare the relative efficiencies between the two methods, so we include a factor for this as well.

\end{itemize}

\section{Comparison of Simulation to Measurements \label{results}}

\begin{figure}
  \begin{center}
  	\includegraphics[width=0.5\textwidth]{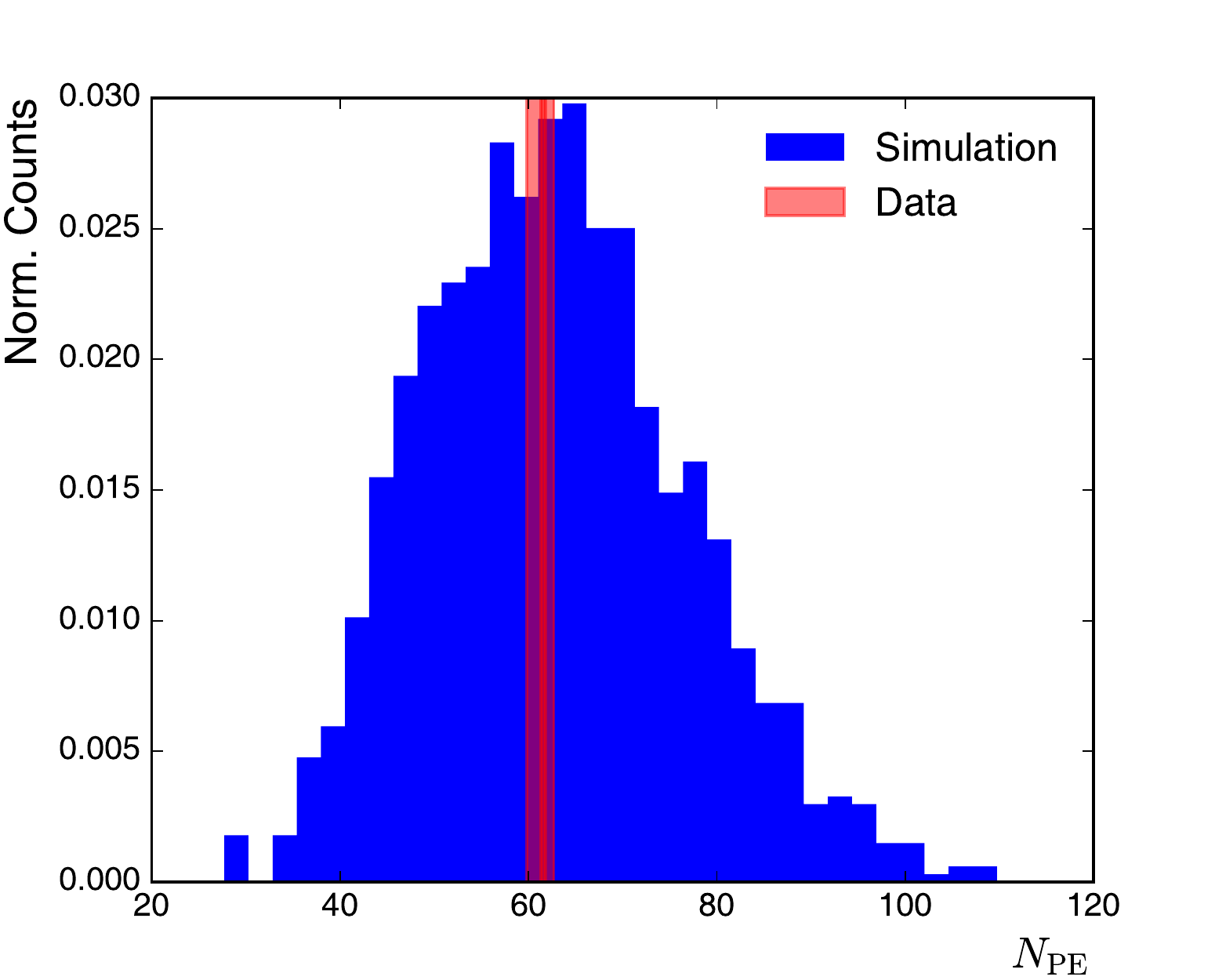}~\includegraphics[width=0.5\textwidth]{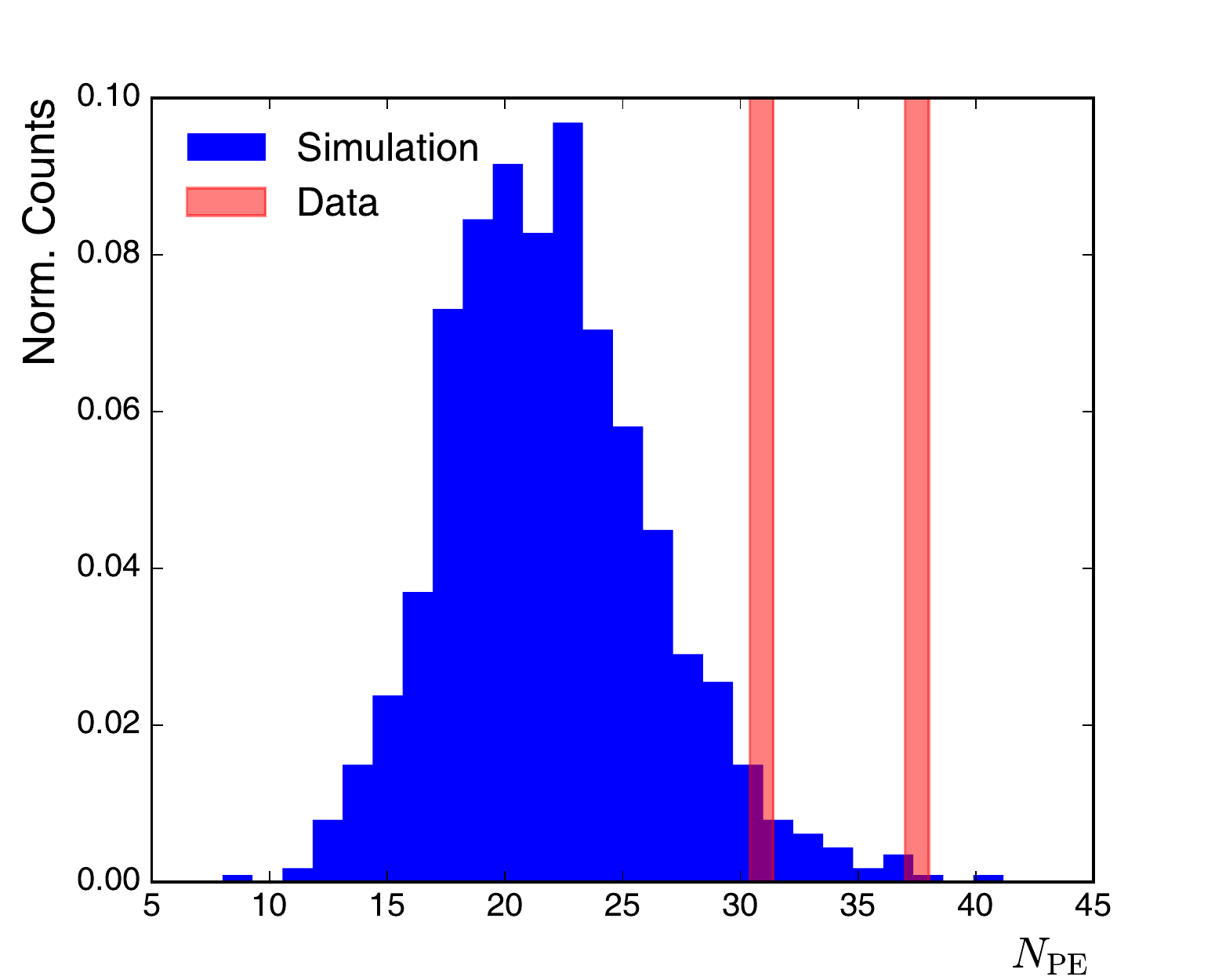}
  \end{center}
\caption{The range of the $N_{\rm PE}$ values predicted by the model 
for the the near configuration (left)
and far configuration (right).   The red lines indicate 
the measured $N_{\rm PE}$ values (the width of the line is the uncertainty) described in section~\ref{sec:absnorm}, table~\ref{NPESPE}; in the left plot, the two measurements overlap.
  \label{NPErange}}
\end{figure}

The simulation of the near and far alpha source measurements
produces a range of $N_{\rm PE}$ values indicating the spread
that can occur given the systematic errors we have input. The distributions for simulated $N_{\rm PE}$ are shown in Fig.~\ref{NPErange}, where the left figure shows the near source position and the right shows the far source position. The red lines on the histograms indicate the four $N_{\rm PE}$ measurements discussed in Section \ref{sec:absnorm}.   
One measurement (far measurement 1) lies in the tail of the distribution, but is
not outside the simulation prediction.  The other three measurements lie well within the predicted region.    

We have noted already in section~\ref{sec:sim} that the radial
dependence of the quantum efficiency, across the TPB plate, predicted by the simulation (figure~\ref{fig:plateeff}) is the same as that seen in the measurements (figure~\ref{results_figure}) described in section~\ref{sec:posdep}, and is Gaussian.  So the simulation is able to reproduce
both the absolute scale of the quantum efficiency and also
the position dependence across the TPB plate.  We are
thus ready to update our mathematical model (equation~\ref{dqe_def}) for the differential
quantum efficiency, $Q(r,\phi)$;
\begin{equation}
Q(r,\phi)=Q_0\exp(-r^2/2\sigma^2)
\end{equation}
where $Q_0$ represents the quantum efficiency at the center of the TPB plate.  The value of $\sigma$ observed in the measurements ranges from 8.8 cm at 1050 V to 11.7 cm at 1350 V; 
the value determined by the ensemble of simulations, which depends on a particular description of the PMT, is $8.00\pm0.05$ cm.  

The value of $Q_0$ is the product of a number of factors, one from the ensemble of simulations and the others from Table~\ref{tab:params}.
\begin{eqnarray*}
Q_0 &=& [\textrm{PMT collection efficiency at center of plate (as in figure~\ref{fig:plateeff})}=0.24\pm0.02]\\
& & \times[\textrm{PMT quantum efficiency at 87 K}=0.153\pm0.008]\\
& & \times[\textrm{TPB re-emission probability}=0.40\pm0.04]\\
& & \times[\textrm{TPB coating factor} = 0.67\pm0.06]\\
& & \times[\textrm{Temperature correction}=1.22\pm0.15]\\
&=& 0.012\pm 0.002
\end{eqnarray*}
Using $Q_0=0.012\pm0.002$ and $\sigma=8.00\pm0.05$ cm, we are able
to generate a value for the global quantum efficiency 
(equation~\ref{dqe_int}) averaged
over the surface of the TPB plate.
\begin{equation}
\left<Q\right> = Q_0 \frac{2}{R^2} \int_0^R dr\, r\,\exp(-r^2/2\sigma^2)=0.0055\pm0.0009
\end{equation}

Future work which would seek to improve the precision of our estimate should aim to prevent or correct systematic uncertainties associated with making measurements with different apparatus configurations.  Changing the configuration from the near to far setup required opening and then refilling the cryostat; this process can introduce possible effects due to measurements at different purity or temperature. An improvement to the setup would be the ability to make measurements at different source positions without disassembling the apparatus.

\section{Conclusion}

We have created a simulation model for the optical units used in the MicroBooNE experiment.  It is able to reproduce the measurements of PMT response using alpha particle scintillation in liquid argon to illuminate the unit, and it also reproduces the measured position dependence of the response across the surface of the TPB plate.  Therefore, this model can be applied to the MicroBooNE optical simulation with confidence. This model can also easily be adapted to other light collection systems, such as the flat-panel systems planned for use in DUNE.

\section{Acknowledgements}
This work was supported by the US Department of Energy, Office of Science, Medium Energy Nuclear Physics program, grant number DE-FG02-94ER40847; by the US National Science Foundation, grant number NSF-PHY-1505855; 
and by Fermi National Accelerator Laboratory, which is operated by the Fermi Research Alliance LLC
under Contract No. DE-AC02-07CH11359 with the US
Department of Energy.
We are grateful for the assistance of Stephen Pordes, Ron Davis, and William Miner at the Proton Assembly Building at Fermi National Accelerator Laboratory.

\bibliography{response_function}{}

\end{document}